\documentclass[conference,a4paper]{IEEEtran}
\usepackage[a-2b,mathxmp]{pdfx}[2024/06/15]

\usepackage{etoolbox}
\newbool{showappendix}

\usepackage[utf8]{inputenc} 
\usepackage[T1]{fontenc}
\usepackage{xcolor} 
\usepackage{url}
\usepackage{ifthen}
\usepackage{cite}
\usepackage[cmex10]{amsmath,mathtools} 
\usepackage{amssymb,bbm,relsize}
\usepackage{comment}
\usepackage{enumerate}

\newtheorem{theorem}{Theorem}
\newtheorem{lemma}{Lemma}
\newtheorem{definition}{Definition}
\newtheorem{proposition}{Proposition}

\newcommand{\N}{\mathbb{N}}     
\newcommand{\Nzero}{\mathbb{N}_{\geq 0}}     
\newcommand{\R}{\mathbb{R}}     
\newcommand{\Rzero}{\mathbb{R}_{\geq 0}}     

\newcommand{\one}{\mathbbm{1}} 
\newcommand{\dd}{\mathrm{d}}  
\newcommand{\Prob}{\mathbb{P}}     
\newcommand{\E}{\mathbb{E}}     
\newcommand{\I}{\mathbb{I}}     
\newcommand{\IR}{\overline{\mathbb{I}}}     
\newcommand{\FT}{\mathcal{FT}}
\newcommand{\dEntropy}{h}

\newcommand{\INcat}{C}   
\newcommand{\incat}{c}   
\newcommand{\IN}[1]{X_{#1}}   
\newcommand{\OUT}[1]{Y_{#1}}   

\newcommand{\elCount}[2][t]{N_{#1}(#2)}   
\newcommand{\margCount}[1][t]{N^{\mathrm{M}}_{#1}}
\newcommand{\pointP}[1]{\tau_{#1}}   
\newcommand{\subpointP}[1]{\check{\tau}_{#1}}   
\newcommand{\margpointP}[1]{\tau^{\mathrm{M}}_{#1}}   
\newcommand{\interocc}[1]{W_{#1}}   
\newcommand{\rDvec}{\overline{\boldsymbol{\lambda}}}   

\newcommand{\sMP}[1]{Z_{#1}}   
\newcommand{\margsMP}[1]{Z^{\mathrm{M}}_{#1}}   
\newcommand{\embMC}[1]{\zeta_{#1}}   
\newcommand{\subembMC}[1]{\check{\zeta}_{#1}}   
\newcommand{\margembMC}[1]{\zeta^{\mathrm{M}}_{#1}}   
\newcommand{\embMCreal}[1]{z_{#1}}   
\newcommand{\margInitVec}{\boldsymbol{\eta}}
\newcommand{\elmargInit}[1]{\eta_{#1}}

\newcommand{\bret}[1]{V_{#1}}

\newcommand{\elembMat}[2]{P_{#1 #2}}

\newcommand{\SMK}{\boldsymbol{Q}}   
\newcommand{\smkd}{\boldsymbol{q}}   
\newcommand{\elSMK}[2]{Q_{#1 #2}}   
\newcommand{\elsmkd}[2]{q_{#1 #2}}   
\newcommand{\subSMK}{\check{\boldsymbol{Q}}}   
\newcommand{\subsmkd}{\check{\boldsymbol{q}}}   
\newcommand{\elsubSMK}[2]{\check{Q}_{#1 #2}}   
\newcommand{\elsubsmkd}[2]{\check{q}_{#1 #2}}   

\newcommand{\margsmkd}{\Tilde{\boldsymbol{q}}}   
\newcommand{\elmargSMK}[2]{\Tilde{Q}_{#1 #2}}   
\newcommand{\elmargsmkd}[2]{\Tilde{q}_{#1 #2}}   

\newcommand{\elSoujD}[1]{G_{#1}}   

\newcommand{\margSoujd}{\Tilde{\boldsymbol{g}}}   
\newcommand{\elmargSoujd}[1]{\Tilde{g}_{#1}}   

\newcommand{\elSoujM}[1]{\mu_{#1}}   
\newcommand{\elInteroccD}[1]{F_{#1}}   
\newcommand{\elInteroccd}[1]{f_{#1}}   
\newcommand{\SurvD}{\boldsymbol{S}}   
\newcommand{\elSurvD}[1]{S_{#1}}   

\newcommand{\elReccM}[1]{\pi_{#1}} 

\newcommand{\stsp}{E}   
\newcommand{\substsp}{\check{E}}   
\newcommand{\margstsp}{E_{\mathrm{M}}}   
\newcommand{\orthstsp}{\hat{E}}   
\newcommand{\idx}{i} 
\newcommand{\idxset}{\mathcal{I}} 
\newcommand{\coarse}{\Pi} 
\newcommand{\margFilt}[1]{\mathcal{F}^{\mathrm{M}}_{#1}}
\newcommand{\SuffStat}[1]{\boldsymbol{\theta}_{#1}}

\mathtoolsset{showonlyrefs=true} 
 \booltrue{showappendix} 

\begin{document}
\title{Mutual Information of a class of Poisson-type Channels using Markov Renewal Theory} 

\author{%
  \IEEEauthorblockN{Maximilian Gehri, Nicolai Engelmann and Heinz Koeppl}
  \IEEEauthorblockA{Dept. of Electrical Engineering, Centre for Synthetic Biology\\
                    Technische Universit{\"{a}}t Darmstadt\\
                    Darmstadt, Germany\\
                    Email: \{maximilian.gehri, nicolai.engelmann, heinz.koeppl\}@tu-darmstadt.de}
}

\maketitle

\begin{abstract}
    The mutual information (MI) of Poisson-type channels has been linked to a filtering problem since the 70s, but its evaluation for specific continuous-time, discrete-state systems remains a demanding task. As an advantage, Markov renewal processes (MrP) retain their renewal property under state space filtering. This offers a way to solve the filtering problem analytically for small systems. We consider a class of communication systems $X \to Y$ that can be derived from an MrP by a custom filtering procedure. For the subclasses, where (i) $Y$ is a renewal process or (ii) $(X,Y)$ belongs to a class of MrPs, we provide an evolution equation for finite transmission duration $T>0$ and limit theorems for $T \to \infty$ that facilitate simulation-free evaluation of the MI $\mathbb{I}(X_{[0,T]}; Y_{[0,T]})$ and its associated mutual information rate (MIR). In other cases, simulation cost is reduced to the marginal system $(X,Y)$ or $Y$. We show that systems with an additional $X$-modulating level $C$, which statically chooses between different processes $X_{[0,T]}(c)$, can naturally be included in our framework, thereby giving an expression for $\mathbb{I}(C; Y_{[0,T]})$. Our primary contribution is to apply the results of classical (Markov renewal) filtering theory in a novel manner to the problem of exactly computing the MI/MIR. The theoretical framework is showcased in an application to bacterial gene expression, where filtering is analytically tractable.
\end{abstract}
\section{Introduction}
The Poisson channel was introduced as a model of the direct-detection photon channel in the context of optical communication \cite{bar1969communication,macchi1972estimation}. Since then it has been discussed for various fields of application, e.g., inter- and intracellular signal transduction \cite{thomas2016capacity,eckford2018channel,sinzger2020poisson} and neuroscience \cite{cannon2016analytical}. In all fields, jump-like models account for the discrete nature of reactions. We motivate our analysis of the MI from a biochemical perspective. There is mounting evidence that cells encode information in the steady state concentration or spatio-temporal variation of signalling molecules \cite{behar2010understanding,purvis2013encoding,friedrich2019stochastic}. Previous publications considered both, channels between trajectories \cite{tostevin2009mutual,lestas2010fundamental,nakahira2018fundamental,duso2019path,sinzger2020poisson,reinhardt2023path,sinzger2023asymptotic,moor2023dynamic} and channels with a static input but dynamic output \cite{selimkhanov2014accurate,cepeda2019estimating,tang2021quantifying}, however mostly not confined to Poisson-type channels. We view the Poisson channel as a prototype to advance the methodology to exactly (up to numerical errors) compute the MI for more general jump-like channels with static and dynamic input. While exact, stochastic simulation-based methods \cite{reinhardt2023path,moor2023dynamic,tang2021quantifying}, analytical approximations \cite{moor2023dynamic,tostevin2009mutual} and approximate simulation-based methods \cite{duso2019path,pasha2012computing} exist, exact analytical or numerical solutions could so far only be obtained for particular low dimensional models \cite{sinzger2020poisson,sinzger2023asymptotic}, in the limit of a discrete-time approximation \cite{thomas2016capacity,thomas2016shannon}, or for the discrete-time Poisson channel \cite{shamai1990capacity}.  Our framework extends the analytical and numerical results to a class of MrPs and facilitates marginal simulation.

\subsection{Definitions and problem statement}
Consider a probability space $(\Omega,\mathcal{F},\Prob)$ with complete filtration $\mathcal{F} = (\mathcal{F}_t)_{t\geq 0}$. Denote the Borel sets on $\Rzero$ by $\mathfrak{B}$. Unless stated otherwise we use the following conventions throughout: Vectors are row vectors, a lower case letter corresponding to a function $\boldsymbol{A}(t)$ symbolizes its density and vice versa, e.g., $\boldsymbol{a}(t) \coloneqq \partial_{t} \boldsymbol{A}(t)$, and for objects defined with indices we use the object without indices for their matrix notation, e.g., $\boldsymbol{A} \coloneqq (A_{i j})_{i\in \mathcal{I}, j \in \mathcal{J}}$ with some countable sets $\mathcal{I}$, $\mathcal{J}$. The MI between two $\mathcal{F}$-adapted stochastic processes on a finite interval $[0,T]$, $T>0$ is defined via the Radon-Nikodym derivative between the joint probability measure $\Prob^{\IN{} \OUT{}}_{T}$ of input and output trajectories $(\IN{[0,T]},\OUT{[0,T]})= (\IN{t},\OUT{t})_{t\in [0,T]}$ and product of the respective marginal measures $\Prob^{\IN{}}_{T}$, $\Prob^{\OUT{}}_{T}$ of $\IN{[0,T]}= (\IN{t})_{t\in [0,T]}$, $\OUT{[0,T]}= (\OUT{t})_{t\in [0,T]}$:
\begin{equation}
    \I(\IN{[0,T]};\OUT{[0,T]}) \coloneqq \E \left[ \ln{\frac{\dd \Prob^{\IN{} \OUT{}}_{T}}{ \dd(\Prob^{\IN{}}_{T} \otimes \Prob^{\OUT{}}_{T} ) }} \right]
    \label{eq:MI-Def}
\end{equation}
The MIR is defined as its asymptotic slope
\begin{equation}
    \IR(\IN{};\OUT{}) \coloneqq \lim_{T \to \infty} \frac{1}{T} \I (\IN{[0,T]};\OUT{[0,T]}),
    \label{eq:MIR-Def}
\end{equation}
with $(\IN{},\OUT{})=(\IN{t},\OUT{t})_{t \in \Rzero}$.
Let $\mathcal{F}^{\IN{} \OUT{}}=(\mathcal{F}^{\IN{} \OUT{}}_t)_{t \geq 0}$ and $\mathcal{F}^{\OUT{}}=(\mathcal{F}^{\OUT{}}_t)_{t \geq 0}$ be the internal filtrations of $(\IN{},\OUT{})$, respectively, $\OUT{}$, where $\mathcal{F}^{\IN{} \OUT{}}_t = \sigma((\IN{s},\OUT{s}), 0 \leq s \leq t)$ and $\mathcal{F}^{\OUT{}}_t = \sigma(\OUT{s}, 0 \leq s \leq t)$. $\mathcal{F}^{\IN{}}$ be analogous. The evaluation of the MI between jump-like processes has long been linked to a filtering problem \cite{Bremaud1972PhD,grigelionis1974mutual,boel1975martingales}. Set $\phi(z) \coloneqq z \ln(z) \, \one_{(0,\infty)}(z)$ with $\one$ denoting the indicator function of the measurable set in its lower index. Let $\OUT{}$ with $\OUT{0}=0$ be a \emph{Poisson-type process} with $\mathcal{F}$-intensity $\lambda$, i.e., $M_{t} \coloneqq \OUT{t} - \int_0^t \lambda_s \, \dd b(s)$ is an $\mathcal{F}$-local martingale, where $\lambda=(\lambda_t)_{t \geq 0}$ is a non-negative $\mathcal{F}$-predictable process and $b:\Rzero \to \Rzero$ is a non-negative right-continuous non-decreasing function \cite{liptser2001statisticsII}. For simplicity we set $b(t)=t$, restricting the Stieltjes-Lebesgue integral to a standard Lebesgue integral. If $\E\left[ \left|  \ln{\frac{\dd \Prob^{\IN{} \OUT{}}_{T}}{ \dd(\Prob^{\IN{}}_{T} \otimes \Prob^{\OUT{}}_{T} ) }} \right| \right]< \infty$, then
\begin{equation}
    \I (\IN{[0,T]};\OUT{[0,T]}) = \int_{0}^{T} \E[\phi(\lambda^{\IN{} \OUT{}}_t)] - \E[ \phi(\lambda^{\OUT{}}_t)] \, \dd t,
    \label{eq:MI-Liptser}
\end{equation}
where $\lambda^{\IN{} \OUT{}}_t = \E[\lambda_t \mid \mathcal{F}^{\IN{} \OUT{}}_{t-}]$ is the $\mathcal{F}^{\IN{} \OUT{}}$-intensity of $Y$ and $\lambda^{\OUT{}}_t= \E[\lambda_t \mid \mathcal{F}^{\OUT{}}_{t-}]$ is the $\mathcal{F}^{\OUT{}}$-intensity of $Y$ \cite{liptser2001statisticsII}. Evaluating these conditional expectations poses a filtering problem. A Poisson-type channel is actually a directed channel by definition, as it is assumed that the input dynamics is independent of the output dynamics, i.e., $\Prob(\IN{t}=x \mid \mathcal{F}^{\IN{} \OUT{}}_s)=\Prob(\IN{t}=x \mid \mathcal{F}^{\IN{}}_s)$ for all $t\geq s\geq 0$. This follows directly from the Randon-Nikodym derivative between general jump-like processes (c.f. \cite{boel1975martingales, duso2019path}). The result is a sum of expressions, resembling the right-hand side of \eqref{eq:MI-Liptser}. Massey \cite{massey1990causality} first resolved the misconception in the historical notion of feedback, as e.g. used in \cite{kabanov1978capacity}, which views feedback as the dependence of $\lambda^{\IN{} \OUT{}}$ on the history of $\OUT{}$, i.e., $\lambda^{\IN{} \OUT{}}_t \neq \E[\lambda_t \mid \mathcal{F}^{\IN{}}_{t-}]$ for some $t\geq 0$. According to Massey, such a self-dependence of the $\OUT{}$-dynamics is better understood as a channel with memory on the past output trajectory. Concluding, if feedback is used, the mutual information does not obey $\eqref{eq:MI-Liptser}$. The presented filtering procedure can, however, be applied to a more general class of channels, e.g., with feedback-use or a more complex output.

For compactness and readability we specify the notation of objects that are used throughout. The reader is referred to \cite{liptser2001statisticsII,bremaud1981point,asmussen2008applied} for more detailed definitions. Let $(\pointP{n})_{n \in \N}$ be an $\mathcal{F}$-adapted non-explosive point process on $[0,\infty]$ with interoccurrence times $\interocc{n+1} \coloneqq (\pointP{n+1}-\pointP{n})$ for all $n \in \Nzero$ with $\pointP{n}<\infty$, and $(\embMC{n})_{n \in \N}$ be a sequence of random variables on a finite or countable state space $\stsp$. The joint sequence $(\embMC{n},\pointP{n})_{n \in \N}$ be an $\stsp$-marked point process. If not explicitly stated otherwise, we will assume that $\pointP{0}=0$, $\embMC{0}=\embMCreal{0} \in \stsp_0$ (some countable space). The $\stsp$-marked point process is equivalently described by the tuple of counting processes $(\elCount[]{z}, z \in \stsp)$ with $\elCount{z} \coloneqq \sum_{n \in \N} \one_{\{\pointP{n} \leq t\}}\one_{\{\embMC{n} = z\}}$ for all $t\geq 0$. The internal filtration of the $\stsp$-marked point process is thus $\mathcal{S}=(\mathcal{S}_t)_{t \geq 0}$ with $\mathcal{S}_t \coloneqq \sigma(\elCount[s]{A}, 0\leq s \leq t, A \subseteq \stsp) = \bigvee_{z \in \stsp} \mathcal{F}^{N}_t(z)$, where $\mathcal{F}^{N}_t(z)=\sigma(\elCount[s]{z}, 0\leq s \leq t)$ describes the internal filtration of $\elCount[]{z}$. Yet another equivalent description is the stochastic process $\sMP{}=(\sMP{t})_{t \geq 0}$ with $\sMP{t} \coloneqq \sum_{n \in \N} \embMC{n} \one_{[\pointP{n}, \pointP{n+1})}(t)$. Note however, that $\mathcal{F}^{\sMP{}}_t =\sigma(\sMP{s}, 0 \leq s \leq t) \subseteq \mathcal{S}_t$ with inequality if $\Prob( \embMC{n+1} = \embMC{n} )>0$ for some $n \in \N$ since $\sMP{}$ does not carry information on self-transitions.

Throughout we assume that $\interocc{n+1}$ conditionally on $\mathcal{S}_{\pointP{n}}= \sigma(\pointP{0}, \embMC{0}, \dots, \pointP{n}, \embMC{n})$ admits an absolutely continuous distribution, such that for every $n \in \Nzero$, $z \in \stsp$ and $A \in \mathfrak{B}$
\begin{equation}
\begin{split}
        &\Prob(\interocc{n+1} \in A, \embMC{n+1}=z \mid \mathcal{S}_{\pointP{n}})(\omega)\\
        =& \elInteroccD{z}^{(n+1)}(A, \omega) = \int_{A} \elInteroccd{z}^{(n+1)}(s,\omega) \dd s
\end{split}
\end{equation}
where $\elInteroccd{z}^{(n+1)} \colon \Rzero \times \Omega \to \Rzero$  is $\mathfrak{B} \otimes \mathcal{S}_{\pointP{n}}$-measurable. Setting $\elInteroccd{}^{(n+1)} \coloneqq \sum_{z \in \stsp} \elInteroccd{z}^{(n+1)}$, this definition implies
\begin{equation}
    \Prob(\interocc{n+1} \in A\mid \mathcal{S}_{\pointP{n}})(\omega)
    = \elInteroccD{}^{(n+1)}(A, \omega) = \int_{A} \elInteroccd{}^{(n+1)}(s,\omega) \dd s.
\end{equation}
Note that $\Prob(\interocc{n+1} = \infty \mid \mathcal{S}_{\pointP{n}}) = 1 - \int_{\Rzero} \elInteroccd{}^{(n+1)}(s) \dd s \geq 0$. Under the above assumptions, Br{\'{e}}maud \cite{Bremaud1972PhD} established the following theorem, which is the basis of this paper.
\begin{theorem}[Regenerative form of the intensity \cite{bremaud1981point}]
For each $z \in \stsp$ $\elCount[]{z}$ is a Poisson process with $\mathcal{S}$-intensity $\lambda(z)$, such that
\begin{equation}
    \lambda_{t}(z) = \!\! \sum_{n \in \Nzero} \!\! \frac{\elInteroccd{z}^{(n+1)}(t - \pointP{n})}{1- \elInteroccD{}^{(n+1)}(t - \pointP{n})} \one_{(\pointP{n},\pointP{n+1}]}(t) \one_{\{\pointP{n} < \infty\}}.
    \label{eq:regenerative-intensity-Bremaud}
\end{equation}
\end{theorem}

This expression facilitates solving the filtering problems in \eqref{eq:MI-Liptser}. Remaining tasks for a particular communication system are then just to derive all $\{\elInteroccd{z}^{(n+1)} \mid z \in \stsp, n \in \Nzero \}$ and the probability of the history $\{(\embMC{n}, \pointP{n}) \mid t < \pointP{n+1} \}$ for all $t \in [0,T]$, and integrate with any tractable method. For Markov renewal systems $(\IN{},\OUT{})$, and renewal $\OUT{}$ the expressions are of particularly simple form and analytical limits are accessible.

We now further restrict $(\embMC{n},\pointP{n})_{n \in \N}$ to be a time-homogeneous MrP, i.e., for all $t\geq 0, z^{\prime} \in \stsp$ it obeys the Markov renewal property
\begin{equation}
\begin{split}
    &\Prob(\embMC{n+1}=z^{\prime}, \pointP{n+1}\leq t \mid \mathcal{S}_{\pointP{n}}) \\
    =& \Prob(\embMC{n+1}=z^{\prime}, \pointP{n+1}\leq t \mid \sigma(\embMC{n}, \pointP{n}))
\end{split}
\label{eq:MrP-property}
\end{equation}
and its \emph{semi-Markov kernel} (sMk) is defined as
\begin{equation}
    \elSMK{z}{z^\prime}(t) \coloneqq \Prob(\embMC{n+1}=z^{\prime}, \interocc{n+1}\leq t \mid \embMC{n} = z)
\end{equation}
for $z,z^{\prime} \in \stsp$ is independent of $n \in \Nzero$. The sMk completely determines the dynamics of the process, analogously to the infinitesimal generator in the case of Markov jump processes. By our previous assumption, the sMk is absolutely continuous with density $\smkd(t)$. For any MrP, $(\embMC{n})_{n \in \N}$ is a Markov chain with \emph{transition probabilities} $P_{z z^{\prime}} \coloneqq \lim_{t \to \infty} \elSMK{z}{z^{\prime}} (t)$ \cite{cinlar1969markov}. The stochastic process $\sMP{}$ is the \emph{semi-Markov process} associated with the MrP. For finite times $\elSoujD{z}(t) \coloneqq \sum_{z^{\prime} \in \stsp} \elSMK{z}{z^{\prime}}(t)$ defines the sojourn time distribution of state $z$. Note however, that $\elSoujD{z}(\infty)\coloneqq 1- \lim_{t \to \infty} \elSoujD{z}(t)$ can be non-vanishing if the state $z$ is partially absorbing and then $\elSoujD{z}(\infty) + \sum_{z^{\prime}} P_{z z^{\prime}} = 1$. As a regularity condition for infinite $\stsp$ we assume $\sup_{z \in \stsp} \elSoujD{z}<1$ for some $t>0$ as in \cite{cinlar1969markov}. 
Note that any irreducible, recurrent MrP is non-explosive. Lastly, $\elSurvD{z} (t) \coloneqq 1 - \elSoujD{z}(t)$ is the \emph{survival function of $z$} and $\bret{}=(\bret{t})_{t \geq 0}$ with $\bret{t} \coloneqq t- \min_{n} \{t-\pointP{n} \mid \pointP{n} \leq t\}$ denotes the \emph{backward-recurrence time} of $(\pointP{n})_{n \in \N}$.

\subsection{Outline}
First, we develop a filtering procedure centered around the use of Anderson's filtering Theorem for MrPs \cite{cinlar1969markov} in Sec. \ref{sec:filtering}. This procedure yields explicit expressions of \eqref{eq:regenerative-intensity-Bremaud} and the probability density of the system history at time $t\geq 0$. If the filtered process is an MrP, then computing $t \mapsto \E[\phi(\lambda_{t}(z))]$ is reduced to solving a system of Volterra integral equations of second kind and integrating over time. In Sec. \ref{sec:LimThm} we derive analytical expressions that reduce the asymptotic filtering problem appearing in \eqref{eq:MIR-Def} to the finite time problem for the same class.
In Sec. \ref{sec:EvalMI} we relate results from previous sections to computing the MI/MIR in different cases, particularly for evaluating $\I (\INcat, \OUT{})$ with an $\IN{}$-modulating static random variable $\INcat{}$. Finally, the theoretical framework is applied to a bacterial gene expression model, where filtering is analytically tractable in Sec. \ref{sec:Application}. The proofs of the propositions and a step-by-step application of the filtering procedure are provided in the appendix \cite{gehri2024mutual}.

\section{Filtering for Markov renewal processes}
\label{sec:filtering}
Here we present a filtering procedure, which is centered around the fact that time-homogeneous Markov renewal property is preserved under projection of the dynamics to a subset of the state space $\substsp \subset \stsp$, relying on the following theorem.

\begin{theorem}[Anderson's filtering theorem for MrPs \cite{cinlar1969markov}]
Let $(\embMC{n},\pointP{n})_{n \in \Nzero}$ be an MrP with sMk $\SMK$. Let $\substsp \subset \stsp$ and $\orthstsp \coloneqq \stsp \setminus \substsp$. For each $\omega \in \Omega$ with $\embMC{0}(\omega) \in \substsp$ define $L_{0}(\omega) \coloneqq 0$, $L_{m+1}(\omega) \coloneqq \inf \{n > L_{m}(\omega) \mid \embMC{n}(\omega) \in \substsp \}$, $\subembMC{m}(\omega) \coloneqq \embMC{L_{m}(\omega)}(\omega)$,  $\subpointP{m} \coloneqq \pointP{L_{m}(\omega)}(\omega)$, for all $m \in \Nzero$. Now partition the sMk into blocks
\begin{equation}
\SMK =
\begin{pmatrix}
    \boldsymbol{A} & \boldsymbol{B} \\ 
    \boldsymbol{C} & \boldsymbol{D}
\end{pmatrix}
\end{equation}
(or subfamilies of functions if $\stsp$ is countably infinite) such that $\boldsymbol{A}(t) \in \Rzero^{\substsp \times \substsp}$ for all $t \in \Rzero$, and all other blocks accordingly.
Then $(\subembMC{n}, \subpointP{n})_{n \in \Nzero}$ is an MrP whose sMk $\subSMK$ is
\begin{equation}
    \subSMK(t) \! = \! \boldsymbol{A}(t) + \! \int_{0}^{t} \! \boldsymbol{B}(\dd u) \! \int_{0}^{t-u} \sum_{k=0}^{\infty} \boldsymbol{D}^{(k)}(\dd s) \, \boldsymbol{C}(t-u-s),
    \label{eq:Anderson}
\end{equation}
 where $D^{(0)}_{\alpha \beta} = \delta_{\alpha \beta} \delta(t)$ ($\alpha, \beta \in \orthstsp$) with Kronecker-/Dirac-Delta and $\boldsymbol{D}^{(k)}(t)=(\boldsymbol{D} \ast \boldsymbol{D}^{(k-1)})(t)$ with the matrix
 convolution $(\boldsymbol{J} \ast \boldsymbol{K})(t) = \int_{[0,t]} \boldsymbol{J}(\dd s) \boldsymbol{K}(t-s)$ for $\boldsymbol{J} \colon \Rzero \to \R^{\mathcal{I} \times \mathcal{J}}$, $\boldsymbol{K} \colon \Rzero \to \R^{\mathcal{J} \times \mathcal{K}}$ and countable sets $\mathcal{I}, \mathcal{J}, \mathcal{K}$.
\end{theorem}

\subsection{Filtering procedure for Markov renewal system models}
\label{subsec:FilteringProcedure}
Let the MrP $(\embMC{n}, \pointP{n})_{n \in \Nzero}$ describe the system of interest. This process need not be $\mathcal{F}^{\IN{} \OUT{}}$-adapted, but $\mathcal{F}^{\IN{} \OUT{}}_t \subseteq \mathcal{S}_t$ for all $t \geq 0$. So the dynamics of the communication system $(\IN{},\OUT{})$ (only source and output processes) is contained within this model. The ability to describe the system of interest by an MrP is a formal limit of our filtering procedure and defines the considered class of Poisson-type channels.

Both $(\IN{},\OUT{})$ and the marginal output $\OUT{}$ constitute a subset of marked transitions in $(\embMC{n}, \pointP{n})_{n \in \Nzero}$. We describe a 3-step procedure to get the marginal dynamics of the respective \emph{marginal system} $(\margembMC{n},\margpointP{n})_{n \in \Nzero}$ with yet to be defined mark-space $\margstsp$: (i) state-space augmentation, (ii) semi-Markov filtering and (iii) coarse-graining.

(i) Let $\idxset \subseteq \Nzero$ be the set of transition class indices. Each possible transition of $(\embMC{n})_{n \in \Nzero}$ (also self-transitions) is assigned to a class index $\idx \in \idxset$. All transitions that are not observable in the trajectories of the marginal system are assigned to the class with index $0$. All observable transitions are assigned to classes with $i>0$. Two transitions are in the same class $i$, if they are indistinguishable in the trajectories of marginal system. For instance, if the marginal system is $\OUT{}$, a univariate counting process, then $| \idxset |\leq 2$. We define the augmented system $(\embMC{n},\idx_n, \pointP{n})_{n \in \Nzero}$ such that $(\idx_n)_{n \in \Nzero}$ is a random sequence on $\idxset$ and $\idx_n=i$ if $\embMC{n}$ was entered via a transition of class $i$.

(ii) Apply \eqref{eq:Anderson} to the augmented MrP with
\begin{equation}
    \substsp = \{(z,\idx) \in \stsp \times \idxset_0 \mid \exists n \in \N \colon \Prob(\embMC{n}=z, \idx_n = \idx)>0 \}
\end{equation}
and $\idxset_0\coloneqq \idxset \setminus \{0\}$.
We obtain the sMk $\subSMK$ of the filtered augmented MrP $(\subembMC{n},\check{\idx}_n,\subpointP{n})_{n \in \Nzero}$, where by construction $(\subpointP{n})_{n \in \N} = (\margpointP{n})_{n \in \N}$. We denote $\margCount \coloneqq \sum_{n \in \N} \one_{\{\margpointP{n} \leq t \}}$.

(iii) Coarse-graining is a problem-specific step. Consider a surjective mapping $\coarse \colon \substsp \to \margstsp$, such that $\coarse((\subembMC{n}, \check{\idx}_n))= \margembMC{n}$ for all $n \in \Nzero$. We call $\coarse$ the \emph{coarse-graining}. If $\coarse$ is also injective, then the marginal system $(\margembMC{n},\margpointP{n})_{n \in \Nzero}$ is an MrP with sMk $\elsubSMK{\coarse^{-1}(\cdot) \,}{\coarse^{-1}(\cdot)}$, otherwise the Markov renewal property of the $\margstsp$-marked point process representation of the marginal system is lost.

\subsection{Intensity and path-probability of the marginal system}
\label{subsec:Filtered-Intensities}
Consider the filtered augmented system $(\subembMC{n},\check{\idx}_n,\margpointP{n})_{n \in \Nzero}$ with sMk $\subSMK$. Let $\margFilt{}$ be the internal filtration of the marginal system with initial side conditions, i.e., $\margFilt{t} \coloneqq \sigma(\embMC{0},\pointP{0}) \vee \mathcal{S}^{\mathrm{M}}_{t}$ and $\embMC{0}=\xi \in \stsp$, $\pointP{0}=0$ ($\Prob$-a.s.). Denote the interoccurrence times of $(\margpointP{n})_{n \in \Nzero}$ by $(\interocc{n})_{n \in \N}$. Further set $\elmargSMK{\alpha}{\beta}({z^{\prime}}, t) \coloneqq \elsubSMK{\alpha}{\beta}(t) \, \one_{\coarse^{-1}(\{z^{\prime}\})}(\beta)$ and $\elmargSoujd{\alpha}(z^{\prime},t) \coloneqq \sum_{\beta \in \substsp} \elmargsmkd{\alpha}{\beta}(z^{\prime},t)$ for $z^{\prime} \in \margstsp$, $\alpha, \beta \in \substsp$.     
We recursively define $\margFilt{\margpointP{n}}$-adapted $[0,1]^{\substsp}$-valued statistics
\begin{equation}
    \SuffStat{n} \coloneqq \frac{\SuffStat{n-1} \, \margsmkd (\margembMC{n},\interocc{n})}{\left\langle \SuffStat{n-1} , \margSoujd (\margembMC{n}, \interocc{n}) \right\rangle} ,
    \label{eq:Recursive-Theta}
\end{equation}
where $\langle \cdot, \cdot \rangle$ denotes the scalar product. For $\margpointP{0}> \pointP{0}$ define
\begin{equation}
    (\SuffStat{0})_{\alpha} \coloneqq \frac{ \one_{\coarse^{-1} (\{\margembMC{0}\})}(\alpha) \elsubsmkd{(\xi,0)}{\alpha}^{0} (\margpointP{0}) }{\tilde{f}_{\xi \margembMC{0}}(\margpointP{0})},
    \label{eq:SuffStatInitial1}
\end{equation} 
 with $\subsmkd^{0}$ and $\tilde{\boldsymbol{f}}$ as defined in Theorem \ref{thm:FilteredIntensity}. For $\margpointP{0} = \pointP{0}$ set
\begin{equation}
    (\SuffStat{0})_{\alpha} \coloneqq \frac{ \one_{\coarse^{-1} (\{\margembMC{0}\})}(\alpha) \; \Prob ((\subembMC{0}, \check{i}_0)= \alpha \mid \embMC{0}= \xi) }{ \sum_{\beta \in \substsp} \one_{\coarse^{-1} (\{\margembMC{0}\})}(\beta) \; \Prob((\subembMC{0}, \check{i}_0)= \beta \mid \embMC{0}= \xi)}.
    \label{eq:SuffStatInitial2}
\end{equation}
 In the case where $\{(\xi^{\prime}, i^{\prime}) \in \substsp \mid \xi^{\prime} = \xi \} \neq \emptyset$, the maximum entropy distribution $\Prob((\subembMC{0}, \check{i}_0)= \beta \mid \embMC{0}= \xi) = \frac{ \delta_{(\xi, i^{\prime}) \beta}}{\sum_{\Tilde{i} }\delta_{(\xi, \Tilde{i}) \beta}}$ is a possible choice. Other distributions are also possible and depend on further model assumptions. Lastly, define
 \begin{equation}
     \Lambda_{z}^{0}(t,\xi) \! \coloneqq \! \frac{\tilde{f}_{\xi z}(t)}{1 \! - \! \int_{0}^{t} \sum_{y \in \margstsp} \tilde{f}_{\xi y}(s) \dd s}, \,
    \Lambda_{z}(t,\SuffStat{}) \! \coloneqq \! \frac{\left\langle \SuffStat{}, \margSoujd(z,t) \right\rangle}{\left\langle \SuffStat{}, \SurvD (t) \right\rangle}
 \end{equation}
with the survival functions $\elSurvD{\alpha}(t)= 1 - \sum_{\beta \in \substsp} \elsubSMK{\alpha}{\beta}(t)$. We call $\Lambda_{z}^{0}$ the \emph{transient hazard function of state $z$} and $\Lambda_{z}$ \emph{recurrent hazard function of state $z$}.
\begin{theorem}
$\margCount[] (z)$ admits the $\margFilt{}$-intensity
\begin{equation}
    \lambda^{\mathrm{M}}_t(z) = \Lambda_{z}^{0}(t-,\xi) \one_{(0,\margpointP{0}]}(t) + \Lambda_{z}(\bret{t-},\SuffStat{\margCount[t-]}) \one_{(\margpointP{0},\infty)}(t),
\end{equation}
  where $\tilde{f}_{\xi z}$ is obtained by a modified filtering procedure, such that for step (ii) apply \eqref{eq:Anderson} with $\substsp^{0} \coloneqq \{(\xi,0)\} \cup \substsp$ to get $\subsmkd^{0}$. Subsequently assign $\tilde{f}_{\xi z}(t) \coloneqq \sum_{\alpha \in \coarse^{-1} (\{z\})} \elsubsmkd{(\xi,0)}{\alpha}^{0} (t)$.
For all $t > 0$ the path-probability is
   \begin{equation}
    \begin{split}
        &p(\margCount=n, \bret{t}=v, \margpointP{n} = t_n ,\margembMC{n}=z_n, \dots \mid \pointP{0}=0, \embMC{0}=\xi) \\
        =& \delta(t-v-t_{n}) \left\langle \SuffStat{0} \margsmkd(z_1,t_{1}-t_{0}) \cdots  \margsmkd(z_n,t_{n}-t_{n-1}),\SurvD(v) \right\rangle.
    \end{split}
   \end{equation}
   
If $\coarse$ is also injective, then $\Lambda_{z}(\bret{t},\SuffStat{\margCount}) = \Lambda_{z}^{\mathrm{MrP}}(\bret{t}, \margsMP{t})$ with
\begin{equation}
    \Lambda_{z}^{\mathrm{MrP}}(t, z^{\prime}) = \elsubsmkd{\coarse^{-1}(z^{\prime})\,}{\coarse^{-1}(z)}/\elSurvD{\coarse^{-1}(z^{\prime})} (t)
    \label{eq:MrP-hazard-function}
\end{equation}
   \label{thm:FilteredIntensity}
\end{theorem}

\subsection{Evolution equation for Markov renewal systems}
\label{subsec:EvolEqMrP}
Let $\sMP{}$ be a semi-Markov marginal system with sMk $\SMK$, $\margFilt{}$-intensities $\lambda (z)$ and initial distribution $\eta_{z} \coloneqq \Prob( \embMC{0}=z)$ for all $z \in \margstsp$.
\begin{proposition}
    For all $z \in \margstsp$ it holds
    \begin{equation}
        \begin{split}
            \E\left[\phi(\lambda_{t}(z)) \right] =& \sum_{y \in \margstsp} \Big[ \eta_{y} \elsmkd{y}{z} (t) \, \ln \left( \Lambda_{z}^{\mathrm{MrP}}(t,y) \right) \\
            &+ \int_{(0,t]} \ln \left( \Lambda_{z}^{\mathrm{MrP}}(v,y) \right) \elsmkd{y}{z} (v) \E [\lambda_{t-v}(y)] \dd v \Big]
        \end{split}
    \end{equation}
    \vspace{-20pt}\\
    and the renewal densities
    obey $(\E[\lambda_{t}(z)])_{z \in \margstsp} \eqqcolon\rDvec (t) =  \margInitVec \smkd (t)  + \int_{(0,t]} \rDvec (s) \smkd (t-s) \dd s$, a Volterra Eq. of 2nd kind \cite{polyanin2008handbook}.
    \label{prop:phi-renewal-density-evolution}
\end{proposition}
\begin{IEEEproof}
    The probability density of $(V_t,Z_t)$ follows by \cite{cinlar1969markov}, Eq. (8.6). Evaluate the expectations $\E [\phi(\Lambda_{z}^{\mathrm{MrP}}(V_t,Z_t))]$ and $\E [\Lambda_{z}^{\mathrm{MrP}}(V_t,Z_t)]$ using the density.
\end{IEEEproof}
\section{Limit theorems}
\label{sec:LimThm}

To compute the MIR via \eqref{eq:MI-Liptser} we are interested in the asymptotic intensities and in particular
\begin{equation}
    \lim_{T \to \infty} \frac{1}{T} \int_{0}^{T} \E \left[ \phi( \Lambda_{z}(\bret{t-},\SuffStat{\margCount[t-]}) ) \right] \dd t .
    \label{eq:limit-Lipster-term-marginal}
\end{equation}
For fully history-dependent marginal systems exist
 numerical methods, that use properties of $(\bret{t},\SuffStat{\margCount})$ to obtain \eqref{eq:limit-Lipster-term-marginal} \cite{sinzger2023asymptotic}. 

In contrast, for semi-Markov marginal systems it is possible to derive \eqref{eq:limit-Lipster-term-marginal} analytically given some convergence criteria.  We proceed with $\sMP{}$ defined as in Sec. \ref{subsec:EvolEqMrP} and denote by $\elReccM{z}=\E \left[ \min \{ \pointP{n} - \pointP{1} \mid \embMC{n} = z \} \mid \embMC{1}=z \right]$ the \emph{mean recurrence time in state $z$}. Further assume that $\sMP{}$ is irreducible, recurrent and aperiodic. Let $\FT[\cdot]$ denote the Fourier transform.

In order to obtain \eqref{eq:limit-Lipster-term-marginal} we use the key renewal theorem \cite{cinlar1969markov} to formulate conditions on $\smkd$, such that $t \mapsto \E[\lambda_{t}(z)]$ converges. For brevity we drop the indices of $\Lambda^{\mathrm{MrP}}$ and $\margstsp$.

\begin{theorem}
    For each $z \in \stsp$ let $\elSMK{y}{z} \neq 0$ only for finitely many $y \in \stsp$. If for all $z,y \in \stsp$ the sMk-density element $\elsmkd{y}{z}$ is (i) directly Riemann integrable, or (ii) $\FT [\elsmkd{y}{z}] \in L^1$, then
    \begin{equation}
        \begin{split}
           & \lim_{T \to \infty} \frac{1}{T} \int_{0}^{T} \E[ \phi(\lambda_{t}(z))] \dd t\\
            =& \sum_{y \in \stsp} \frac{1}{\elReccM{y}} \elembMat{y}{z} (\ln (\elembMat{y}{z}) + \dEntropy(\sigma_{y z}) - \E[ \ln(\elSurvD{z}(\sigma_{y z})) ] ) < \infty,
            \label{eq:MIR-dEntropy-MrP}
        \end{split}
    \end{equation}
    where $L^1$ denotes the integrable functions, $\dEntropy$ is the differential entropy and $\sigma_{y z} \sim \elSMK{y}{z}/\elembMat{y}{z}$ is the holding time in $y$ given a transition to $z$. In particular, if $|\stsp|=1$, i.e., $(\pointP{n})_{n \in \Nzero}$ is a renewal process, then
    \begin{equation}
        \lim_{T \to \infty} \frac{1}{T} \int_{0}^{T} \E[ \phi(\lambda_{t})] \dd t = \frac{1- \dEntropy(\tau)}{\E[\tau]},
        \label{eq:MIR-dEntropy-renewal}
    \end{equation}
    where $\tau \sim \elSMK{}{}$ is the holding time of the process.
    \label{thm:MrP-Limits}
\end{theorem}
For an account of direct Riemann integrability (d.R.i.) and verification for a particular function, see e.g. \cite{asmussen2008applied,iksanov2016renewal}. For instance, any exponential probability density is d.R.i. Further, any non-negative affine combination $t \mapsto \sum_{j} a_j p_{j}(t) \geq 0$, $a_j \in \R$, of d.R.i. probability density functions $p_{j} \colon \Rzero \to \Rzero$ is itself d.R.i. Such affine combinations arise via filtering of Markov systems, as in our example (Sec. \ref{sec:Application}).

\section{Mutual information}
\label{sec:EvalMI}
We now gather our results in order to evaluate the MI or the MIR. For the Poisson channel the marginal system corresponding to $\OUT{}$ satisfies $|\margstsp|=1$. We distinguish between the case (i) that $\OUT{}$ is a renewal process, and case (ii) that $\OUT{}$ is generally history-dependent. In case (i) we may use Proposition \ref{prop:phi-renewal-density-evolution} to evaluate the $\lambda^{\OUT{}}$-related term in \eqref{eq:MI-Liptser}. For the MIR we use \eqref{eq:MIR-dEntropy-renewal}. In case (ii) we use Theorem \ref{thm:FilteredIntensity}. We may use the path-probabilities and the intensity representation to express the term in \eqref{eq:MI-Liptser} as a one-dimensional integral over a series of integrals over $[0,t]^n$, $n \in \N$, whose simulation-free evaluation may be expensive. Another method uses Monte-Carlo integration over the path space as outlined below.

Similarly, we can distinguish the cases, where the marginal point process representing $(\IN{},\OUT{})$ is either an MrP or generally history dependent. We call $(\margembMC{n},\margpointP{n})_{n \in \Nzero}$ a \emph{representation} of $(\IN{},\OUT{})$ if $\margFilt{t} = \mathcal{F}_{t}^{\IN{} \OUT{}}$ for all $t \geq 0$.
For the MrP case we may again use Proposition \ref{prop:phi-renewal-density-evolution} and \eqref{eq:MIR-dEntropy-MrP} to evaluate the $\lambda^{\IN{}\OUT{}}$-related term in \eqref{eq:MI-Liptser}.

In the case that $X$ has an absorbing state the methodology of the above cases remains, but $\OUT{}$ is typically a process that exhibits a transient part as long as $\IN{}$ is in a transient state.

Finally, consider the case of an additional modulation of $\IN{}$ by a static random variable $\INcat$ on the state space $\mathcal{\INcat}$. First note that \eqref{eq:MI-Def} and \eqref{eq:MI-Liptser} still apply in both cases, $\I(\INcat;\OUT{[0,T]})$ and $\I(\IN{[0,T]}(\INcat);\OUT{[0,T]})$, if $\INcat$ is interpreted as a jump-process that has a random initial value, but is otherwise constant. To see this, note that $\INcat \to (\IN{}(\INcat),\OUT{}(\INcat))$ is a Markov chain of length 1, where $\incat$ determines the sMk $\SMK^{(\incat)}$ of the model of each subsystem $(\IN{}(\incat),\OUT{}(\incat))$ and the initial condition $\embMC{0}(\incat)$. The sMk of the full model can hence be written as the block-diagonal matrix $\SMK \coloneqq \mathrm{diag}(\SMK^{(\incat)} \colon \incat \in \mathcal{\INcat})$ on the state space $\mathcal{\INcat} \times \stsp$ with lexicographic order. We write $\embMC{0}(\INcat)=(\embMC{0}, \INcat)$ as a pair of independent random variables and fix $\embMC{0}= \xi \in \stsp$. The filtered augmented state space generalizes to $\mathcal{\INcat} \times \substsp$, where steps (i) and (ii) apply to each $\incat$ independently, i.e., $\subSMK \coloneqq \mathrm{diag}(\subSMK^{(\incat)} \colon \incat \in \mathcal{\INcat})$.
In the case $\margpointP{0}=\pointP{0}$, we obtain the $\mathcal{F}^{\INcat\OUT{}}$-intensity $\lambda^{\INcat \OUT{}}$, respectively, the $\mathcal{F}^{\OUT{}}$-intensity $\lambda^{\OUT{}}$ under modulation of $\INcat$, via Theorem \ref{thm:FilteredIntensity} by evaluating $\SuffStat{0}$ with
\bgroup \setlength{\abovedisplayskip}{5pt} \setlength{\belowdisplayskip}{5pt} \setlength{\abovedisplayshortskip}{5pt} \setlength{\belowdisplayshortskip}{5pt}
\begin{align}
    &\Prob((\INcat, \subembMC{0}, \check{i}_0)= \beta \mid \INcat=\incat, \embMC{0}= \xi) = \frac{ \delta_{(\incat,\xi, i^{\prime}) \beta}}{\sum_{\Tilde{i} }\delta_{(\incat, \xi, \Tilde{i}) \beta}} \label{eq:SuffStat0-Cconditional}\\
    &\Prob((\INcat, \subembMC{0}, \check{i}_0)= \beta \mid \embMC{0}= \xi) \\
    =& \sum_{\incat \in \mathcal{\INcat}} \Prob((\INcat, \subembMC{0}, \check{i}_0)= \beta \mid \INcat=\incat, \embMC{0}= \xi) \Prob(\INcat=\incat),
    \label{eq:SuffStat0-C}
\end{align}
\egroup
with maximum entropy assumption used in \eqref{eq:SuffStat0-Cconditional}.

Finally, by virtue of \cite{liptser2001statisticsII} (p. 288) and identifying $\pointP{n}(z)$ with the $n$-the jump time of $N(z)$, we obtain
\bgroup \setlength{\abovedisplayskip}{5pt} \setlength{\belowdisplayskip}{5pt} \setlength{\abovedisplayshortskip}{5pt} \setlength{\belowdisplayshortskip}{5pt}
\begin{align*}
        &\int_{0}^{T} \E \left[ \phi( \lambda_{t}(z) ) \right] \dd t
        = \E \left[ \int_{0}^{T}  \ln( \Lambda_{z}(\bret{t-},\SuffStat{\margCount[t-]})) \dd N_t(z)  \right] \\
        =& \E \Big[ \sum_{\substack{n \in \N \\ \pointP{n}(z) \leq T}}  \ln(\Lambda_{z}( \interocc{m(n)},\SuffStat{m(n)-1}))  \Big] \; \text{ with } \pointP{m(n)} = \pointP{n}(z)
\end{align*}
\egroup
(and analogously for $\Lambda^{\mathrm{MrP}}$) for use with simulations.
\section{Application -- Bacterial Promoter Regulation}
\label{sec:Application}

\begin{figure}
  \includegraphics[width=\columnwidth]{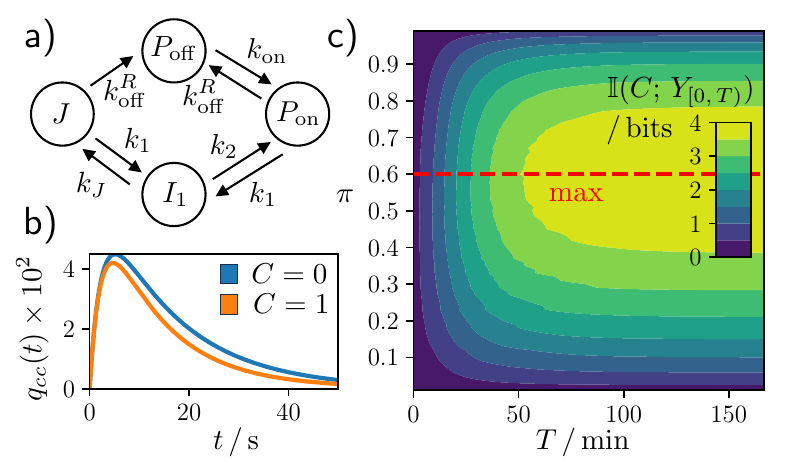}
  \caption{The synthetic biology scenario; \emph{a)} shows the transition graph of the model of interest. Transitions into $J$ are equivalent to arrivals at $\OUT{}$. State $J$ mirrors state $P_{\mathrm{on}}$ in terms of outgoing transitions, such that being in state $J$ is equivalent to being in $P_{\mathrm{on}}$. Transitions into $P_{\mathrm{off}}$ model off-switching of $\IN{}$ and transitions $P_{\mathrm{off}} \to P_{\mathrm{on}}$ model on-switching; \emph{b)} analytically obtained inter-arrival time densities of $\OUT{}$ for both outcomes of $\INcat{}$; \emph{c)} $\I\left(\INcat{};\,\OUT{\left[0,\,T\right]}\right)$ for different success probabilities $\pi$ and different $T$ for $10^5$ trajectories of $\OUT{}$. The dashed red line highlights the value of $\pi$ for  attained maximal $\I\left(\INcat{};\,\OUT{\left[0,\,T_{\max}\right]}\right)$. The kinetic parameters are listed in section \ref{sec:Application}.}
  \label{fig:figure_synbio}
\end{figure}

As a use-case, we consider a scenario of bacterial gene expression with a promoter regulated by a repressor in the domain of synthetic biology. We investigate the case of an extended two-state active-inactive Markovian promoter model with additional intermediate states as motivated in \cite{hakkinen2012evolving}. In many cases, transcription regulation in bacteria can be modeled by a reversible reaction that represents repressor binding and unbinding events \cite{sanchez2011effect}. In this case, the binding rate $k_{\text{off}}^{R}\equiv k_{\text{off}}[R]$ of a repressor $R$ to its binding site is the product of the intracellular repressor concentration $\left[R\right]\in\mathbb{R}_{\geq0}$ and the association constant $k_{\text{off}}>0$. The dissociation constant $k_{\text{on}}>0$ models its unbinding rate.
We include one intermediate rate-limiting step during transcription, so that our reaction network is given as
\[
P_{\text{off}}\xrightleftharpoons[k_{\text{off}}^{R}]{k_{\text{on}}}P_{\text{on}}\xrightleftharpoons[k_{2}]{k_{1}}I_{1}\stackrel{k_{J}}{\rightarrow}P_{\text{on}}+\mathrm{mRNA},
\]
where $P_{\text{off}}$ denotes the promoter's inactive state due to a bound repressor $R$, and $P_{\text{on}}$ denotes the promoter's active state. The active state $P_{\text{on}}$ can transition to the intermediate state $I_{1}$ in a reversible reaction with rate $k_{1}$ towards $I_{1}$ and rate $k_{2}$ “back” to $P_{\text{on}}$. The modulated birth reaction of the mRNA is irreversible, happens with rate $k_{J}$ and reverts the promoter back to its active state $P_{\text{on}}$. The transition graph of a finite state representation of this system is shown in Fig. \ref{fig:figure_synbio}a. We identify $\IN{t}=1$ with $\sMP{t} \in \{ P_{\text{on}}, I_1, J \}$, $\IN{t}=0$ with $\sMP{t} = P_{\text{off}}$ and $\OUT{} = N(\mathrm{J})$. Several extrinsic conditions \cite{alradhawi2019multimodality} motivate the existence of a probability distribution over the concentration of repressors $[R]$ in the intracellular environment.

Under idealistic conditions, we analyze the mutual information $\I\left(\INcat{};\,\OUT{\left[0,\,T\right]}\right)$ between the counting process of produced mRNA $\OUT{}$
and a Bernoulli variable $\INcat{}\,:\,\Omega\rightarrow\left\{ 0,\,1\right\}$ with success probability $\pi\in\left[0,\,1\right]$ that dictates the repressor concentration $[R]$. Thus, we set $[R]\coloneqq \INcat{}[R]_1 + (1-\INcat{})[R]_0$, where $[R]_{0}$ and $[R]_{1}\neq[R]_{0}$ are two fixed concentrations. As $[R]$ is time-invariant, the $\mathcal{F}^{\INcat{}\OUT{}}$-adapted process $(C, \OUT{})$ has a renewal property and application of \eqref{eq:Anderson} allows us to obtain an analytical solution for the distribution $\elSMK{c}{c}$ of its inter-arrival times (c.f. Fig \ref{fig:figure_synbio}b). The $\mathcal{F}^{\OUT{}}$-adapted intensity of the non-Markovian counting process $\OUT{}$ reduces to
\begin{equation}
    \lambda^{\OUT{}}_{t} = \frac{\sum_{\incat} p(\incat) \elsmkd{\incat}{\incat}(W_1) \cdots \elsmkd{\incat}{\incat}(W_{\OUT{t-}}) \elsmkd{\incat}{\incat}(\bret{t-})}{\sum_{\incat} p(\incat) \elsmkd{\incat}{\incat}(W_1) \cdots \elsmkd{\incat}{\incat}(W_{\OUT{t-}}) \elSurvD{\incat}(\bret{t-})},
\end{equation}
by the discussion at the end of Sec. \ref{sec:EvalMI}, where $\elSurvD{\incat} \coloneqq 1 - \elSMK{c}{c}$ is the survival function of the holding times of $\OUT{}$ given $\INcat=\incat$. We calculate $\I\left(\INcat{};\,\OUT{\left[0,\,T\right]}\right)$ using Monte Carlo integration for several configurations of $\pi$ and time-limits $T\in[0,\,T_{\max}]$. In Fig. \ref{fig:figure_synbio}c we show a contour plot of $\I\left(\INcat{};\,\OUT{\left[0,\,T\right]}\right)$ over $T$ and $\pi$. The dashed red line shows the attained maximal $\I\left(\INcat{};\,\OUT{\left[0,\,T_{\max}\right]}\right)$ at around $\pi\approx0.6$. The capacity-achieving distribution is asymmetric, as has been observed in other cases of modulated Poisson channels \cite{davis1980capacity, kabanov1978capacity}.
The kinetic constants used have been chosen to align with those in \cite{sanchez2011effect}. These are $k_{\mathrm{on}}=0.0023\mathrm{s}^{-1}$, $k_{\mathrm{off}}=0.0027\mathrm{s}^{-1}\mathrm{nM}^{-1}$ and for simplicity, 
$k_{1}=k_{2}=k_{J}=0.165\mathrm{s}^{-1}$. We chose the two repressor concentrations to be $[R]_0=1\mathrm{nM}$ and $[R]_1=10\mathrm{nM}$ w.r.t. a normalized volume to model a fold-change of $10$.

\section{Conclusion}
We considered models with Markov renewal description and used a filtering procedure around Anderson's filtering theorem to get expressions for the $\mathcal{F}^{\IN{} \OUT{}}$- 
and $\mathcal{F}^{\OUT{}}$-intensity of $\OUT{}$. The filtering procedure involves a series of matrix convolutions, which is intractable for larger systems. For small systems, particularly if the marginal process $\OUT{}$ is a renewal process, we derived evolution equations and limit theorems that can be used to exactly evaluate the MIR and the MI for finite transmission duration. In particular, for a class of systems we established a link between the terms of the MIR and the differential entropy of the holding times of $(X,Y)$ and $Y$. For systems with a tractable filtering problem, marginal simulation is facilitated via Theorem \ref{thm:FilteredIntensity}. We presented our method with a synthetic biology motivated channel. More applications of our framework to the various fields of interest remain to be discussed. Proofs and further details are provided in \cite{gehri2024mutual}.


\section*{Acknowledgment}
We thank Mark Sinzger-D'Angelo for helpful discussions.


\ifbool{showappendix}{
\IEEEtriggeratref{21}
}{\IEEEtriggeratref{18}}

\bibliographystyle{IEEEtran}
\bibliography{Bibliography/biblio_math,Bibliography/biblio_IT,Bibliography/biblio_bio}



\ifbool{showappendix}{
\newpage
\appendix
\subsection{Notes on practical issues}
In this section we discuss four issues of practical relevance: 1) benefits of the Laplace transform for solving the filtering problem and the Volterra Eq. of the renewal density vector, 2) how to construct the sMk, 3) we elaborate on our notion of a representation of a stochastic process and how different representations may exhibit different properties, and 4) we state some propositions to check the conditions of the limit theorems, i.e., if a function is d.R.i. or has an integrable Fourier transform.

\subsubsection{Laplace transform and filtering}
The use of \eqref{eq:Anderson} in Anderson's filtering theorem seems unpractical due to the series of higher-order self-convolutions. However, since here $\SMK$ is absolutely continuous, the component-wise Laplace transform of $\smkd$ exists and \eqref{eq:Anderson} has an equivalent Laplace transformed version (denoted with upper asterisk)
\begin{equation}
    \subsmkd^{\ast}(s) = \boldsymbol{a}^{\ast}(s) + \boldsymbol{b}^{\ast}(s) \sum_{k=0}^{\infty} (\boldsymbol{d}^{\ast}(s))^k \, \boldsymbol{c}^{\ast}(s), \quad s \in \mathbb{C}.
    \label{eq:Anderson-Laplace}
\end{equation}
By the convolution theorem for the Laplace transform, the series of self-convolutions reduces to a geometric series after the Laplace transform. Similarly, the Volterra integral Eq. in Proposition \ref{prop:phi-renewal-density-evolution} reduces to the algebraic equation
\begin{equation}
    \rDvec^{\ast}(s) (\one - \smkd^{\ast}(s)) = \margInitVec \smkd^{\ast}(s) , \quad s \in \mathbb{C},
\end{equation}
where $\one$ denotes the identity matrix. Solving the algebraic equation by standard methods decouples the components of $\rDvec^{\ast}(s)$ and subsequent numerical Laplace inversion \cite{hassanzadeh2007comparison} may hence improve the accuracy of a numerical solution to the Volterra Eq. \cite{press2007numerical} (p. 933). It is well known that for $|\stsp|< \infty$ the solution is unique \cite{burton2005volterra} (p. 25).

\subsubsection{Construction of the semi-Markov kernel}
There are several types of random mechanisms to generate sMk density elements $\elsmkd{y}{z}$. We describe two prevalent mechanisms \cite{nunn1977semi} and convert the generator of a Markov jump process to its corresponding sMk. Denote $\Tilde{F}_{y z} (t) \coloneqq \E [\interocc{n+1} \leq t \mid \elembMat{n+1}=z, \elembMat{n}=y] = \frac{\elSMK{y}{z}(t)}{\elembMat{y}{z}}$.

1. After having entered state $y$, first the successor state $l$ is drawn randomly according to the transition probabilities $\elembMat{y}{z}$. Secondly, the holding tim in state $y$, given that the next state is $z$ is drawn according to $\Tilde{F}_{y z}$. Then
\[\elsmkd{y}{z}(t) = \elembMat{y}{z} \Tilde{f}_{y z} (t).\]

2. After having entered state $y$, the holding times $S_{y z}$ for all possible subsequent states $z \in \stsp$ are drawn independently according to the absolutely continuous distribution function $F_{y z}$. The smallest draw determines the successor state and the holding time. If $(s_{y x} \geq 0, x \in \stsp)$ are the drawn realizations, then the successor state is $z = \mathrm{argmin} \{ s_{y x} \geq 0, x \in \stsp \}$ and the holding time is $s_{y z} = \min \{ s_{y x} \geq 0, x \in \stsp \}$. Hence $\elSMK{y}{z}(t) = \Prob (S_{y z} \leq t, S_{y z} < S_{y x} \, \forall x \neq z)$. In this case it can be shown that \[\elsmkd{y}{z} (t) = f_{y z} (t) \prod_{x \neq z} (1 - F_{y x})(t). \]

For a Markov jump process $\sMP{}$ with generator $\Lambda(y, z) = \lim_{s \searrow 0}\frac{1}{s} \Prob(\sMP{t+s}=z \mid \sMP{t}=y)$ for all $z \neq y$ and arbitrary $t\geq 0$, we have \[ \elsmkd{y}{z} (t) = \Lambda(y, z) \, e^{- t \sum_{x \neq y} \Lambda(y, x) }. \]

\subsubsection{Representations of a stochastic process}
In section \ref{sec:EvalMI} we introduced the notion of a representation of a stochastic process. We generalize the notion as follows.
\begin{definition}
Let $\stsp, \Tilde{\stsp}$ be countable or finite sets. $(\embMC{n},\pointP{n})_{n \in \Nzero}$ be an $\stsp$-marked point process with $\pointP{0}=0$ and internal filtration $\mathcal{S}$, where $\mathcal{S}_t \coloneqq \sigma(\embMC{0})\vee \sigma(\elCount[s]{A} \colon 0\leq s \leq t, A \subseteq \stsp)$ and $(N(z) \colon z \in \stsp)$ denotes its equivalent counting process. $X=(X_{t})_{t \in \Rzero}$ be a right-continuous stochastic process on $\Tilde{\stsp}$ with internal filtration $\mathcal{F}^{X}$. We call $(\embMC{n},\pointP{n})_{n \in \Nzero}$ and $X$ \emph{representations} of each other if $\mathcal{S}_t=\mathcal{F}^{X}_t$ for all $t \in \Rzero$. Analogously, two right-continuous stochastic processes (or two marked point processes) are representations of each other, if they have equal internal filtrations.
\label{def:representations}
\end{definition}
Definition \ref{def:representations} establishes an equivalence relation, such that each stochastic processes in the associated equivalence class can be used to express the conditional expectation $\E[\cdot \mid \mathcal{F}^{X}_t]$ as a regular conditional expectation $\E[\cdot \mid X_{[0,t]}= x_{[0,t]}]$.

Note that the trajectory space $\mathbb{D}([0,T],\Tilde{\stsp}) \coloneqq \{ f: [0,T] \to \Tilde{\stsp}, \text{ right-continuous (with left limits)} \}$ together with the Skorokhod metric is a Polish space \cite{billingsley2013convergence}. Using the associated Borel $\sigma$-algebra, \cite{klenke2020probability}, Theorem 8.36, then asserts that the trajectory space is Borel isomorphic to a measurable subset of $\R$. Analogously, $\mathbb{D}([0,T],\Nzero^{\stsp})$ is also a Borel space. The factorization lemma \cite{klenke2020probability} then asserts the existence of measurable maps $\psi_{t} \colon \mathbb{D}([0,t],\Tilde{\stsp}) \to \Nzero^{\stsp}$ for $t>0$, $\psi_0 \colon \Tilde{\stsp} \to \stsp$ and $\Tilde{\psi}_{t} \colon \stsp \times \mathbb{D}([0,t],\Nzero^{\stsp}) \to \Tilde{\stsp}$, such that
\begin{equation}
    \begin{split}
    (N_t(z) \colon z \in \stsp) =& \psi_{t} (X_t \colon t \in [0,t]), \; \psi_0(X_0)=\embMC{0},\\
        X_t =& \Tilde{\psi}_t (\embMC{0}, N_{t}(z) \colon z \in \Tilde{\stsp}, t\in [0,t])
    \end{split}
\end{equation} for all $t\geq0$.
Hence exists a bijective map between the trajectories $X_{[0,t]}$ and $(\embMC{0}, N_{[0,t]}(z) \colon z \in \stsp)$ for all $t\in \Rzero$.
Conversely, by the factorization lemma, the existence of the families of functions $(\psi_0, \psi_t \colon t >0)$ and $(\Tilde{\psi}_t \colon t \in \Rzero)$ asserts that $\mathcal{S}_t=\mathcal{F}^{X}_t$ for all $t \in \Rzero$, which can be used to show that two processes are representations of each other.

Our filtering framework leverages the use of different representations of a stochastic process. In particular, some processes on a countably infinite $\Tilde{\stsp}$ can be represented as processes with finite $\stsp$, thereby reducing the dimensionality of the filtering problem.
We exemplify different properties of distinct representations by considering a simple gene expression model with a leaky two-state promoter.

\begin{figure}
    \centering
    \includegraphics{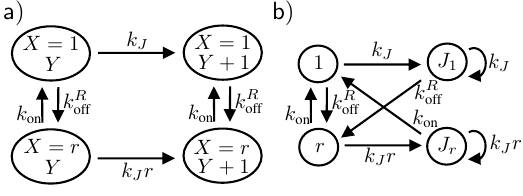}
    \caption{Representations of simple gene expression model with a leaky two-state promoter as the input $\IN{}$, where $\IN{}=1$ denotes the active state and $\IN{}=r \in (0,1)$ is the leaky inactive state with $r$ being the fraction of activity in the inactive state. The output $\OUT{}$ models mRNA synthesis events. a) models the output directly as a counting process such that the state space is infinite, while b) represents the model on a finite state space with two jump states, $J_1$ and $J_r$.}
    \label{fig:leakage-model}
\end{figure}
Fig. \ref{fig:leakage-model} depicts two representations of the reaction network
\[
Y + (X=r) \stackrel{k_{J} r}{\leftarrow}(X=r)\xrightleftharpoons[k_{\text{off}}^{R}]{k_{\text{on}}}(X=1)\stackrel{k_{J}}{\rightarrow}(X=1) + Y.
\]
Fig. \ref{fig:leakage-model} a) shows the state transition diagram of this reaction network on the state space $\Tilde{\stsp} \coloneqq \{r,1\} \times \Nzero$. The model is directly represented as the joint process $(\IN{},\OUT{})$ of input and output and thus no filtering or coarse graining is required to obtain the intensity process $\lambda^{\IN{} \OUT{}}_t = k_{J} \IN{t}$. Since $(\IN{},\OUT{})$ is a Markov process, it also satisfies the Markov renewal property. However, it is not irreducible, such that the limit theorems in \ref{sec:LimThm} do not apply. However, since the input $\IN{}$ is conditionally independent of the output $\OUT{}$, we may still analytically evaluate
\begin{equation}
    \begin{split}
        \E[\phi(\lambda^{\IN{} \OUT{}}_t)] =& k_J r \ln (k_J r) \Prob(\IN{t}=r) \\
        &+ k_J \ln (k_J) \Prob(\IN{t}=1),\\
        \lim_{T \to \infty} \frac{1}{T} \int_{0}^{T}\E[\phi(\lambda^{\IN{} \OUT{}}_t)] \, \dd t =&  k_J r \ln (k_J r) \lim_{t \to \infty} \Prob(\IN{t}=r) \\
        & + k_J \ln (k_J) \lim_{t \to \infty}\Prob(\IN{t}=1).
    \end{split}
\end{equation}

In contrast, Fig. \ref{fig:leakage-model} b) depicts a finite state space representation $(\embMC{n},\pointP{n})_{n \in \Nzero}$ on $\stsp \coloneqq \{r,1,J_1,J_r\}$, which is an irreducible and recurrent Markov process. Let $(\tau^{\IN{} \OUT{}}_n)_{n \in \Nzero}$ denote the jump times of $(\IN{},\OUT{})$ and $(N_{t})_{t \in \Rzero}$ the counting process of all jumps of the $\stsp$-marked point process. The two processes are representations of each other as the relations
\begin{equation}
    \begin{split}
        \OUT{t} =& N_{t}(J_r) + N_{t}(J_1)\\
        \IN{t} =& \one_{\{1, J_1\}}(\embMC{N_t}) + r \, \one_{\{r, J_r\}}(\embMC{N_t}), \quad \embMC{0} =  \IN{0}\\
        N_{t}(1) =& \sum_{n=1}^{\infty} \one_{\{1\}}(X_{\tau^{\IN{} \OUT{}}_n}), \quad N_{t}(r) = \sum_{n=1}^{\infty} \one_{\{r\}}(X_{\tau^{\IN{} \OUT{}}_n})\\
        N_{t}(J_1) =& \int_{0}^{\infty} \one_{\{1\}}(X_{t}) \, \dd Y_t, \quad N_{t}(J_r) = \int_{0}^{\infty} \one_{\{r\}}(X_{t}) \, \dd Y_t,
    \end{split}
\end{equation}
 show. The above relations yield $\lambda^{\IN{} \OUT{}}_{t} = \lambda_{t}(J_1) + \lambda_{t}(J_r) = k_{J} \left( \one_{\{1, J_{1}\}}(Z_t) + r \one_{\{r, J_{r}\}}(Z_t) \right)$, where $Z_t$ denotes the associated semi-Markov process. Thus $\E[\phi(\lambda^{\IN{} \OUT{}}_t)]$ can be evaluated via the mass function $z \mapsto \Prob(Z_{t}= z)$, $z \in \stsp$.

Yet another representation of $(\IN{},\OUT{})$ is found by applying the coarse-graining with $\coarse(J_r)=\coarse(J_1)=J$, $\coarse(r)=r$ and $\coarse(1)=1$. We obtain $N_{t}(J)=\OUT{t}$, $\IN{t} = \margembMC{r(t)}$ with $r(t) \coloneqq \mathrm{argmin}_{n} (\margpointP{n} \leq t, \margembMC{n} \in \{r,1\})$, and $N(1)$, $N(r)$ as in the relations for representation b). The resulting process on $\margstsp \coloneqq \{r,1,J\}$ is irreducible and recurrent, but it does no longer satisfy the Markov renewal property. So while we can obtain the intensity $\lambda^{\IN{} \OUT{}}$ via Theorem \ref{thm:FilteredIntensity}, neither the MrP evolution equation nor the limit theorems apply.

To obtain a representation of the marginal output $\OUT{}$ we may apply the filtering step (ii) with $\substsp \coloneqq \{ J_1, J_r \}$ and then use the coarse-graining $\coarse$. Then $\OUT{t}=N^{M}_{t}(J)$. Importantly, the $\substsp$-marked point process (before the coarse-graining) is not a representation of $\OUT{}$ since the processes $N_t(J_1)$, $N_t(J_r)$ cannot be expressed as a function of $\OUT{[0,t]}$.

\subsubsection{On the conditions of the limit theorems}
Here we discuss some simple complementary propositions that simplify showing that a function is d.R.i or has an integrable Fourier transform. In the context of d.R.i. functions we refer the reader to \cite{asmussen2008applied} (p. 154) and \cite{iksanov2016renewal} (pp. 213). The propositions therein are complemented with the following proposition.
\begin{proposition}
    \begin{enumerate}[(a)]
        \item Let $f \colon \Rzero \to \Rzero$ be d.R.i. and $r \colon \Rzero \to \Rzero$ be a probability density function. Then $f \ast r$ is d.R.i.
        \item Any non-negative linear combination $t \mapsto \sum_{j} a_j p_{j}(t) \geq 0$, $a_j \in \R$, of d.R.i. probability density functions $p_{j} \colon \Rzero \to \Rzero$ is itself d.R.i.
    \end{enumerate}
\end{proposition}

\begin{IEEEproof}
    \begin{enumerate}[(a)]
        \item Apply \cite{iksanov2016renewal}, Lemma 6.2.1 (c).
        \item We use the notation in \cite{asmussen2008applied}. Let $a,b \in \R$ and $h>0$ such that condition (ii) in \cite{asmussen2008applied}, Proposition 4.1 is satisfied for both d.R.i. functions $f,r$. Then
        \begin{align*}
            &\int_{0}^{\infty} \overline{(a f(x) + b r(x))}_h \, \dd x \\
            \leq & \int_{0}^{\infty} (|a| \overline{f}_{h} (x) + |b| \overline{r}_{h} (x) ) \, \dd x \\
            =& |a| \int_{0}^{\infty} \overline{f}_{h} (x) \, \dd x + |b| \int_{0}^{\infty} \overline{r}_{h} (x) \, \dd x < \infty.
        \end{align*}
        This argument is repeated iteratively.
    \end{enumerate}
\end{IEEEproof}
Note that any exponential probability density function is d.R.i. since it is non-increasing and integrable. The density of a Gamma distribution $t \mapsto \frac{\beta^{\alpha}}{\Gamma(\alpha)} t^{\alpha -1} e^{- \beta t}$ with $\alpha, \beta>0$ is also d.R.i. by \cite{iksanov2016renewal}, Lemma 6.2.1 (b).

In the context of Fourier transforms,the following proposition may be useful. The proof is straighforward and will thus be omitted.
\begin{proposition}
    If $f,g\colon \R \to \Rzero$ are (sub)probability density functions and $\FT[g] \in L^1$, then $\FT[g \ast f] \in L^1$. If $r \in [0,1]$ and additionally $\FT[f] \in L^1$, then $\FT[r g + (1-r) f] \in L^1$.
\end{proposition}
\subsection{MIR of a class of 3-state Markov renewal models}
\label{subsec:MIR-3-state}
For any 3-state model with the marginal state transition diagram as displayed in Fig. \ref{fig:3-state-marginal}, the MIR has the form
\begin{equation}
\begin{split}
    &\IR(\IN{}; \, \OUT{})\\
    =& \frac{1}{\elReccM{\mathrm{J}}} \Big( \dEntropy(\tau) + \elembMat{\mathrm{J}}{\mathrm{J}} (\ln(\elembMat{\mathrm{J}}{\mathrm{J}}) - \dEntropy(\sigma_{\mathrm{J} \mathrm{J}})) \\
    &+ \elembMat{\mathrm{J}}{\, \mathrm{OFF}}\E[ \ln(\elSurvD{\mathrm{J}}(\sigma_{\mathrm{J} \, \mathrm{OFF}})) ] \Big) \\
    &+ \frac{1}{\elReccM{\mathrm{ON}}} \Big( \elembMat{\mathrm{ON}}{\, \mathrm{J}} (\ln(\elembMat{\mathrm{ON}}{\, \mathrm{J}}) - \dEntropy(\sigma_{\mathrm{ON} \, \mathrm{J}})) \\
    &+ \elembMat{\mathrm{ON}}{\, \mathrm{OFF}}\E[ \ln(\elSurvD{\mathrm{ON}}(\sigma_{\mathrm{ON} \, \mathrm{OFF}})) ] + 1 \Big),
\end{split}
\end{equation}
where we adapted the notation in Theorem \ref{thm:MrP-Limits}. Since the transition matrix of the embedded Markov chain is right stochastic, it is sufficiently described by two parameters and thus has invariant measure $\boldsymbol{\alpha} \coloneqq (\elembMat{\mathrm{ON}}{\, \mathrm{J}}, \elembMat{\mathrm{J}}{\, \mathrm{OFF}}, \elembMat{\mathrm{J}}{\, \mathrm{OFF}})$, where $\elembMat{\mathrm{ON}}{\, \mathrm{OFF}} = 1- \elembMat{\mathrm{ON}}{\, \mathrm{J}}$ and $\elembMat{\mathrm{J}}{\, \mathrm{J}} =1 - \elembMat{\mathrm{J}}{\, \mathrm{OFF}}$. Hence by \cite{cinlar1969markov} (Lemma 6.7, Theorem 6.12)
\begin{equation}
    \frac{1}{\elReccM{z}} = \frac{ \elembMat{\mathrm{ON}}{\, \mathrm{J}} \, \one_{\{\mathrm{J}\}} (z) + \elembMat{\mathrm{J}}{\, \mathrm{OFF}} \, \one_{\{\mathrm{ON}, \mathrm{OFF}\}} (z)}{\elembMat{\mathrm{ON}}{\, \mathrm{J}} \, \mu_{\mathrm{J}} + \elembMat{\mathrm{J}}{\, \mathrm{OFF}} \, (\mu_{\mathrm{ON}} + \mu_{\mathrm{OFF}})}.
\end{equation}
Using $\elReccM{\mathrm{J}}= \E[\tau]$ and the identity $\frac{1}{\elReccM{\mathrm{J}}} + \frac{1}{\elReccM{\mathrm{ON}}} = \frac{1}{\elReccM{\mathrm{J}} \elembMat{\mathrm{J}}{\mathrm{J}}}$ we can rewrite the MIR as
\begin{equation}
\begin{split}
    \IR(\IN{}; \, \OUT{}) = \frac{1}{\E[\tau]} \Big( & \ln(\elembMat{\mathrm{J}}{\mathrm{J}}) + \dEntropy(\tau) - \dEntropy(\sigma_{\mathrm{J} \mathrm{J}}) \\
    &+ \frac{\elembMat{\mathrm{ON},}{\mathrm{OFF}}}{\elembMat{\mathrm{J}}{\mathrm{J}}} (1 + \E[ \ln(\elSurvD{\mathrm{ON}}(\sigma_{\mathrm{ON}, \mathrm{OFF}})) ])  \Big).
\end{split}
\end{equation}
Note that $\elSMK{\mathrm{OFF}}{\, \mathrm{ON}}$ is the only sMk element that is not explicitly contained in the above expressions. The time spent in OFF is only implicitly contained in both MIR expressions via $\tau$ and $\E[\tau]$. In a kind of truism, we can interpret this to indicate that the OFF intervals do not contribute to the exchange of information. Note that the example in Sec. \ref{sec:Application} belongs to this class of models.
\begin{figure}
\centering
  \includegraphics[width=0.6\columnwidth]{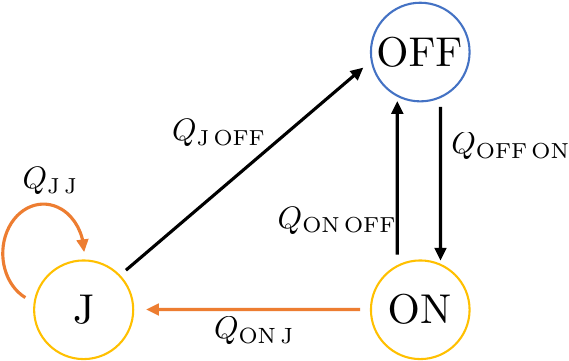}
  \caption{State transition diagram of a class of semi-Markov processes representing the communication system $(\IN{}, \OUT{})$. Being in one of the yellow states means $\IN{t}=1$, whereas being in a blue state means $\IN{t}=0$. Transitions into $\mathrm{J}$ (orange edges) account for arrivals at $\OUT{}$. The marginal process $\OUT{}$ is a renewal process for this class.}
  \label{fig:3-state-marginal}
\end{figure}

\subsection{Proofs}
\begin{IEEEproof}[Proof of Thm. \ref{thm:FilteredIntensity}]
Apply \eqref{eq:regenerative-intensity-Bremaud} to the marked point process $(\embMC{0},\pointP{0},\margembMC{0},\margpointP{0}, \margembMC{1},\margpointP{1}, \dots)$. 
By construction of the filtering procedure for the transient case we obtain $\tilde{F}_{\xi z}(t) = \Prob (\margpointP{0}-\pointP{0}\leq t, \margembMC{0}=z \mid \embMC{0}=\xi)$. After this zeroth transition the sojourn times of the process are those derived from the sMk $\SMK$ by the filtering procedure, where only observable transitions contribute. Set $f^{(n+1)}(z_{n+1},v; t_n, z_n, \dots,t_0,z_{0}, \xi) \coloneqq \partial_{v} \Prob(W_{n+1}\leq v,\margembMC{n+1}=z_{n+1} \mid \margpointP{n}=t_n, \margembMC{n}=z_n, \dots, \margpointP{0}=t_0, \margembMC{0}=z_{0}, \pointP{0}=0, \embMC{0}=\xi)$. Note that
\begin{equation}
    \begin{split}
        &\prod_{m=0}^{n} f^{(m)}(z_{m},t_m -t_{m-1}; t_{m-1}, z_{m-1}, \dots,t_0,z_{0}, \xi) \\
        =& p(\margpointP{n} = t_n ,\margembMC{n}=z_n, \dots \mid \pointP{0}=0, \embMC{0}=\xi)
    \end{split}
    \label{eq:Proof-filtering-product-formula}
\end{equation}
Extending with $(\subembMC{n}, \check{i}_{n})$, $(\subembMC{n-1}, \check{i}_{n-1})$ yields
\begin{equation}
    \begin{split}
        &f^{(n)}(z_{n},t_n -t_{n-1}; t_{n-1}, z_{n-1}, \dots,t_0,z_{0}, \xi) \\
        =& \sum_{\alpha, \beta \in \substsp} \one_{\coarse^{-1}(\{z_{n}\})}(\beta) \elsmkd{\alpha}{\beta}(t_{n}-t_{n-1})\\
        & \cdot \Prob((\subembMC{n-1}, \check{i}_{n-1}) = \alpha \mid \margpointP{n-1}=t_{n-1}, \margembMC{n-1}=z_{n-1}, \dots)\\
        =& \sum_{\alpha \in \substsp} \elmargSoujd{\alpha}(z_{n}, t_{n}-t_{n-1})\\
        & \cdot \Prob((\subembMC{n-1}, \check{i}_{n-1}) = \alpha \mid \margpointP{n-1}=t_{n-1}, \margembMC{n-1}=z_{n-1}, \dots)
    \end{split}
\end{equation}
Apply Bayes rule and extend with $(\subembMC{n-1}, \check{i}_{n-1})$:
\begin{equation}
    \begin{split}
        &\Prob((\subembMC{n}, \check{i}_{n}) = \beta \mid \margpointP{n}=t_{n}, \margembMC{n}=z_{n}, \dots) \\
        =& \Big(\sum_{\alpha \in \substsp} \elmargsmkd{\alpha}{\beta}(z_{n},t_{n}-t_{n-1})\\
        & \cdot \Prob((\subembMC{n-1}, \check{i}_{n-1}) = \alpha \mid t_{n-1}, z_{n-1}, \dots) \Big)\\
        & \cdot \Big( f^{(n-1)}(z_{n-1},t_{n-1} -t_{n-2}; t_{n-2}, z_{n-2}, \dots,t_0,z_{0}, \xi) \Big)^{-1}\\
        =& \Big( \sum_{\alpha \in \substsp} \elmargsmkd{\alpha}{\beta}(z_{n},t_{n}-t_{n-1})\\
        & \, \cdot\Prob((\subembMC{n-1}, \check{i}_{n-1}) = \alpha \mid t_{n-1}, z_{n-1}, \dots ) \Big)\\
        & \cdot \Big( \sum_{\gamma \in \substsp} \elmargSoujd{\gamma}(z_{n-1},t_{n-1}-t_{n-2})\\
        & \, \cdot \Prob((\subembMC{n-1}, \check{i}_{n-1}) = \gamma \mid t_{n-1}, z_{n-1}, \dots) \Big)^{-1}
    \end{split}
\end{equation}
Identifying $\SuffStat{n} = (\Prob((\subembMC{n}, \check{i}_{n}) = \alpha \mid \margFilt{\margpointP{n}}))_{\alpha \in \substsp}$ establishes the recursive relation in Eq. \eqref{eq:Recursive-Theta}. $\SuffStat{0}$ follows analogously. Further identify
\begin{equation}
    f^{(n)}(z,v; \margpointP{n-1} \margembMC{n-1}, \dots) = \left\langle \SuffStat{n-1} , \margSoujd (z, v) \right\rangle
    \label{eq:Proof-filtering-scalar-product-formula}
\end{equation}
Then $\Lambda_{z}$ follows directly by \eqref{eq:regenerative-intensity-Bremaud} and we further obtain 
\begin{equation}
    \begin{split}
        &p(\margCount=n, \bret{t}=v, \margpointP{n} = t_n ,\margembMC{n}=z_n, \dots \mid \pointP{0}=0, \embMC{0}=\xi)\\
        =& p(\margpointP{n+1}>t_n+v, \margpointP{n} = t_n ,\margembMC{n}=z_n, \dots \mid \pointP{0}=0, \embMC{0}=\xi) \\
        & \cdot \delta(t-v-t_{n})\\
        =& \delta(t-v-t_{n}) \, (1- p(\interocc{n+1}\leq v \mid t_{n}, z_{n}, \dots)) \\
        & \cdot p(\margpointP{n} = t_n ,\margembMC{n}=z_n, \dots \mid \pointP{0}=0, \embMC{0}=\xi)\\
        =& \delta(t-v-t_{n}) \sum_{\alpha, \beta \in \substsp} \elSurvD{\beta}(v) \elmargsmkd{\alpha}{\beta}(z_{n},t_{n}-t_{n-1}) \\
        & \cdot p((\subembMC{n-1}, \check{i}_{n-1} ) = \alpha, \margpointP{n-1} = t_{n-1} ,\margembMC{n}=z_{n-1}, \dots \\ & \qquad\qquad\qquad\qquad\qquad\qquad\qquad\qquad \hdots \mid \pointP{0}=0, \embMC{0}=\xi) \\
        =& \delta(t-v-t_{n}) \left\langle \SuffStat{n-1} \margsmkd(z_n,t_{n}-t_{n-1}),\SurvD(v) \right\rangle \\
        &\cdot \prod_{m=1}^{n-1} \left\langle \SuffStat{m-1} ,\margSoujd(z_{m}, t_{m}-t_{m-1}) \right\rangle\\
        =& \delta(t-v-t_{n}) \left\langle \SuffStat{0} \margsmkd(z_1,t_{1}-t_{0}) \cdots  \margsmkd(z_n,t_{n}-t_{n-1}),\SurvD(v) \right\rangle,
    \end{split}
\end{equation}
where in the second last line we used \eqref{eq:Proof-filtering-product-formula}, \eqref{eq:Proof-filtering-scalar-product-formula}, and in the last line iteratively applied the recursive definition of $\SuffStat{}$ in the expression $\left\langle \SuffStat{n-1} \margsmkd(z_n,t_{n}-t_{n-1}),\SurvD(v) \right\rangle$.
\end{IEEEproof}

The following Lemma may be used to conduct the proof of Proposition \ref{prop:phi-renewal-density-evolution}.
\begin{lemma}[Markov renewal equations \cite{cinlar1969markov}]
\begin{equation}
    \begin{split}
        &\Prob(\sMP{t} = z, \bret{t} \leq v) \\
        =& \int_{[0,v]} \one_{[0,t]}(u) \elSurvD{z}(u) \sum_{y \in \stsp} \Prob(\embMC{0}=y) m_{y z}(t-u) \, \dd u \\
        =& \elmargInit{z} \elSurvD{z}(t) + \int_{(0,v]} \one_{[0,t]}(u) \elSurvD{z}(u) \E [ \lambda_{t-u}(z)] \, \dd u
    \end{split}
\end{equation}
with $m_{y z^{\prime}}(t) \coloneqq \delta(t) \delta_{y z^{\prime}} + \E[\lambda_{t}(z^{\prime}) \mid \embMC{0}=y]$.
The renewal density vector obeys $\rDvec (t) =  \margInitVec \smkd (t)  + \int_{(0,t]} \rDvec (s) \smkd (t-s) \dd s$, a Volterra Eq. of second kind \cite{polyanin2008handbook}.
\label{lemma:MrEqs}
\end{lemma}
Note that $\lim_{t \searrow 0}\rDvec (t)$ does not always exist, but $ \lim_{t \searrow 0}\sum_{y} \eta_{z} \elsmkd{y}{z} (t) = \E [\lambda_{0}(z)]$ if the left side exists.

\begin{IEEEproof}[Proof of Lemma \ref{lemma:MrEqs}]
    The evolution equation for $\E[\elCount{z} \mid \embMC{0}=y]$ has a unique solution given by \cite{cinlar1969markov}, Eq. (2.28), via subtraction of the identity matrix. We obtain $\E[\elCount{z} \mid \embMC{0}=y] = \sum_{z^{\prime} \in \stsp} \int_{[0,t]} m_{y z^{\prime}}(u) \elSMK{z^{\prime}}{z}(t-u) \dd u$ with $m_{y z^{\prime}}(t) = \delta(t) \delta_{y z^{\prime}} + \partial_{t} \E[\elCount{z^{\prime}} \mid \embMC{0}=y]$.
    The identity for $m_{y, z^{\prime}}$ follows by \cite{bremaud1981point}, p.27, Eq. (3.2) and Fubini. For the probability evolution equation consider \cite{cinlar1969markov}, Eq. (8.6), plug in $m_{y z^{\prime}}(t)$ and use $\Prob(\sMP{t} = z, \bret{t} \leq v) = \Prob(\sMP{t} = z, \bret{t} \geq 0) - \Prob(\sMP{t} = z, \bret{t} \geq v)$.
\end{IEEEproof}
 
 \begin{IEEEproof}[Proof of Thm. \ref{thm:MrP-Limits}]
     The MrP is aperiodic (or non-lattice) by the the absolute continuity of the sMk density elements \cite{ccinlar1974periodicity,asmussen2008applied}. Further, we have $\elReccM{y}> \elSoujM{y}>0$ for all $y \in \stsp$. Thus for any $z^{\prime}, y, z \in \stsp$ we obtain
     \begin{equation}
         \begin{split}
             &\lim_{t \to \infty}\int_{[0,t)} \elsmkd{y}{z}(t-v) m_{z^{\prime} y}(v) \, \dd v \\
             =& \lim_{t \to \infty}\int_{(0,t)} \elsmkd{y}{z}(t-v) \E[\lambda_{v}(y) \mid \embMC{0}=z^{\prime}]  \, \dd v
             =\frac{\elembMat{y}{z}}{\elReccM{y}} 
         \end{split}
     \end{equation}
 if $\elsmkd{y}{z}$ satifies condition (i) \cite{cinlar1969markov} (Theorem 6.3) or condition (ii) \cite{smith1962necessary} (Lemma 7). The respective sMk density element is essentially bounded, continuous a.e. and satisfies $\lim_{t \to \infty} \elsmkd{y}{z}(t) = 0$ by \cite{asmussen2008applied} (Prop. 4.1 (i)) and \cite{rudin1987complex} (Theorem 9.6). Denote $A_z \coloneqq \{ y \in \stsp \mid \elSMK{y}{z} \neq 0 \}$. Then, by the renewal density evolution equation in Proposition \ref{prop:phi-renewal-density-evolution}, we obtain
 \begin{align*}
     &\E[\lambda_{t}(z)] - \sum_{y \in \stsp} \eta_y \elsmkd{y}{z} (t) = \sum_{y \in A_{z}} \int_{0}^{t} \elsmkd{y}{z}(t-v) \E[ \lambda_{v}(y) ] \, \dd v\\
     \leq & \sum_{y \in A_{z}} \int_{0}^{t} \mathrm{ess \, sup}_{u \geq 0}\elsmkd{y}{z}(u) \E[ \lambda_{v}(y) ] \, \dd v\\
     =& \sum_{y \in A_{z}} \mathrm{ess \, sup}_{u \geq 0}\elsmkd{y}{z}(u) \, \E[ N_{t}(y) ] < \infty
 \end{align*}
 for all $t \geq 0$, i.e., $\E[\lambda_{t}(z)]$ is bounded on finite intervals. Thus the following limit exists and we obtain
  \begin{equation}
        \lim_{t \to \infty} \E [ \lambda_{t}(z) ] = \lim_{t \to \infty} \sum_{y \in A_z} \int_{0}^{t} \E[ \lambda_{v}(y)] \elsmkd{y}{z}(t-v) \, \dd v = \sum_{y \in \stsp} \frac{\elembMat{y}{z}}{\elReccM{y}} .
 \end{equation}
Since the embedded Markov chain is irreducible and recurrent, we obtain
 \begin{equation}
     \lim_{t \to \infty} \E [ \lambda_{t}(z) ] = \sum_{y \in \stsp} \frac{\elembMat{y}{z}}{\elReccM{y}} = \frac{1}{\elReccM{z}} < \infty
 \end{equation}
 by \cite{cinlar1969markov} (Lemma 6.7, Theorem 6.12). Hence we have $\mathrm{sup}_{t \geq 0} \E[\lambda_{t} (z)] < \infty$ and therefore $ \Lambda_z(\bret{t},\sMP{t}) < \infty$ ($\Prob$-a.s.) a.e. (including at infinity). But then $|\phi (\Lambda_z (\bret{t},\sMP{t}) )| < L$ ($\Prob$-a.s.) a.e. for some $L>0$. With the convergence of the renewal densities we reestablish the standard result $(\bret{t}, \sMP{t}) \overset{d}{\rightarrow} (\bret{\infty}, \sMP{\infty})$ (covergence in distribution) and
 \begin{align*}
    &\Prob(\bret{\infty} \leq v, \sMP{\infty} = y) = \lim_{t \to \infty} \Prob(\bret{t} \leq v, \sMP{t} = y)\\
    =& \int_{0}^{v} \elSurvD{y}(u) \lim_{t \to \infty} \E[ \lambda_{t}(y)] \, \dd u = \frac{1}{\elReccM{y}} \int_{0}^{v} \elSurvD{y}(u) \, \dd u 
 \end{align*}
by virtue of the renewal equation of the distribution \cite{cinlar1969markov} and dominated convergence. Since the sMk density is continuous a.e., so is $t \mapsto \Lambda_{z}(t,y)$ for all $y,z \in \stsp$. Thus the continuous mapping theorem implies $\Lambda_{z}(\bret{t}, \sMP{t}) \overset{d}{\rightarrow} \Lambda_{z}(\bret{\infty}, \sMP{\infty})$ \cite{klenke2020probability} (p. 287). Dominated convergence yields
\begin{align*}
&\lim_{t \to \infty} \E[\phi (\Lambda_z (\bret{t},\sMP{t}) )] \\
=& \lim_{t \to \infty} \sum_{y \in \stsp} \int_{0}^{\infty} \phi (\Lambda_z (v,y) )  p(\bret{t} = v, \sMP{t} = y) \, \dd v \\
=& \sum_{y \in A_z} \int_{0}^{\infty} \phi (\Lambda_z (v,y) )  \lim_{t \to \infty} p(\bret{t} = v, \sMP{t} = y) \, \dd v \\
=& \sum_{y \in A_z} \frac{1}{\elReccM{y}} \int_{0}^{\infty} \phi (\Lambda_z (v,y) ) \elSurvD{y}(v)  \, \dd v \\
=& \sum_{y \in \stsp} \frac{1}{\elReccM{y}} \int_{0}^{\infty} \ln (\Lambda_z (v,y) ) \elsmkd{y}{z}(v)  \, \dd v
\end{align*}
and by l'Hospital's rule we obtain
\begin{equation}
    \lim_{T \to \infty} \frac{1}{T}\! \int_{0}^{T} \! \E[ \phi(\lambda_{t}(z))] \dd t
    =\! \sum_{y \in \stsp} \frac{1}{\elReccM{y}} \! \int_{0}^{\infty}\! \ln ( \Lambda_{z}(v,y)) \elsmkd{y}{z}(t) \dd v.
\end{equation}
The r.h.s. of the above Eq. is then rewritten with
    \begin{align*}
        & \int_{0}^{\infty} \ln (\Lambda_{z}(v,y)) \elsmkd{y}{z}(v) \, \dd v\\
        =& \elembMat{y}{z} \int_{0}^{\infty} \frac{\elsmkd{y}{z}(v)}{\elembMat{y}{z}} \left( \ln(\elembMat{y}{z}) + \ln \left( \frac{\elsmkd{y}{z}(v)}{\elembMat{y}{z}} \right) - \ln (\elSurvD{y} (v)) \right) \, \dd v\\
        =& \elembMat{y}{z} \left( \ln(\elembMat{y}{z}) - \dEntropy(\sigma_{y z}) - \E[ \ln (\elSurvD{y} (\sigma_{y z}))] \right),
    \end{align*}
    where we used $\ln (\Lambda_{z}(v,y)) = \ln(\elembMat{y}{z}) + \ln \left( \frac{\elsmkd{y}{z}(v)}{\elembMat{y}{z}} \right) - \ln (\elSurvD{y} (v))$ and $\sigma_{y z} \sim \elSMK{y}{z}/\elembMat{y}{z}$.
     Finally, \eqref{eq:MIR-dEntropy-renewal} follows from \eqref{eq:MIR-dEntropy-MrP} by virtue of the probability integral transform, which states that $\elSurvD{}(\tau) = 1-\elSMK{}{}(\tau) \sim \mathrm{Unif}([0,1])$ and hence \[ \E [\ln (\elSurvD{}(\tau))] = \int_{0}^{1} \ln(t) \, \dd t = -1. \]
 \end{IEEEproof}

\subsection{Simulation-based evaluation}
 Assume that we represented $(\IN{}, \OUT{})$ by an $\stsp$-marked point process such that there is a single state $\mathrm{J} \in \stsp$ with $\OUT{}=N(\mathrm{J})$. For a simulation-based evaluation of the MI we make use of the Eq. at the end of Sec. \ref{sec:EvalMI} and rewrite the r.h.s. of \eqref{eq:MI-Liptser} by
\begin{align*}
    \E \Big[ \sum_{\substack{k \in \N \\ \pointP{k}(\mathrm{J}) \leq T}} & \ln(\Lambda_{\mathrm{J}}^{\IN{} \OUT{}}( \pointP{m(k)}- \pointP{m(k)-1},\SuffStat{m(k)-1}^{\IN{} \OUT{}}))\\
    &- \ln(\Lambda_{\mathrm{J}}^{\OUT{}}( \pointP{k}(\mathrm{J})- \pointP{k-1}(\mathrm{J}),\SuffStat{k-1}^{\OUT{}})) \Big]
\end{align*}
where $m(k)$ is defined such that $\pointP{m(k)} = \pointP{k}(\mathrm{J})$ for all $k\in \N$. Further, $\Lambda_{\mathrm{J}}^{\IN{} \OUT{}}$ and $\SuffStat{}^{\IN{} \OUT{}}$ are defined with respect to the marginal state space $\stsp$, and $\Lambda_{\mathrm{J}}^{\OUT{}}$ and $\SuffStat{}^{\OUT{}}$ with respect to the marginal state space $\{ \mathrm{J} \}$ as described in Sec. \ref{subsec:Filtered-Intensities}. The expectation is evaluated approximately via the law of large numbers. For the numerical evaluation of the intensities at different time points we used the log-sum-exp trick to ensure numerical stability.

\subsection{Application scenario}
Our example system, as displayed in Fig. \ref{fig:figure_synbio} a), reduces to the marginal representation of the communication system $(\IN{}, \OUT{})$ as displayed in Fig. \ref{fig:3-state-marginal}. In the following we apply the filtering procedure outlined in Sec. \ref{subsec:FilteringProcedure} to the system in Fig. \ref{fig:figure_synbio} a), such that the marginal system is a low-dimensional representation of $(\IN{}, \OUT{})$. The sMk of this model with state space $\{\mathrm{J}, P_{\mathrm{on}}, P_{\mathrm{off}}, I_{1} \}$ is
\begin{equation}
    \smkd(t) =
    \begin{pmatrix}
        0 & 0 & k_{\mathrm{off}}^{R} e^{- u_{1} t} &  k_{1} e^{- u_{1} t} \\
        0 & 0 &  k_{\mathrm{off}}^{R} e^{- u_{1} t} &  k_{1} e^{- u_{1} t} \\
        0 & k_{\mathrm{on}} e^{- k_{\mathrm{on}} t} & 0 & 0 \\
        k_{\mathrm{J}} e^{- u_{2} t} & k_{2} e^{- u_{2} t} & 0 & 0
    \end{pmatrix}
\end{equation}
with exit rates $u_{1} \coloneqq k_{\mathrm{off}}^{R} + k_{1}$ and $u_{2} \coloneqq k_{\mathrm{J}} + k_{2}$. We now consider transition class indices $\idxset = \{0,1,2,3\}$, where $0$ are unobservable transitions (all transitions into and out of state $I_1$), $1$ symbolizes ON-switching of the input (i.e., $P_{\mathrm{off}} \to P_{\mathrm{on}}$), $2$ stands for OFF-switching (all transitions into $P_{\mathrm{off}}$), and $3$ are arrivals of $Y$ (all transitions into $J$). The assignment is visualized in Fig. \ref{fig:Filter-Case-Study} a). The augmented state space, following step (i) of the filtering procedure is $\{ (\mathrm{J},3), (P_{\mathrm{on}},1), (P_{\mathrm{off}},2), (P_{\mathrm{on}},0), (I_{1},0) \}$, as in Fig. \ref{fig:Filter-Case-Study} b), and step (ii) yields $\substsp = \{ (\mathrm{J},3), (P_{\mathrm{on}},1), (P_{\mathrm{off}},2) \}$. The Laplace transform of the sMk density on the augmented state space decomposes as
\begin{IEEEeqnarray*}{lllll}
    \boldsymbol{a}^{\ast}(s) =&
    \begin{pmatrix}
        0 & 0 & \frac{k_{\mathrm{off}}^{R}}{u_{1} + s} \\
        0 & 0 & \frac{k_{\mathrm{off}}^{R}}{u_{1} + s}\\
        0 & \frac{k_{\mathrm{on}}}{k_{\mathrm{on}} + s} & 0
    \end{pmatrix},
    & \quad &\boldsymbol{b}^{\ast}(s) =&
    \begin{pmatrix}
        0 & \frac{k_{1}}{u_{1} + s} \\
        0 & \frac{k_{1}}{u_{1} + s} \\
        0 & 0
    \end{pmatrix},\\
    \boldsymbol{c}^{\ast}(s) =&
    \begin{pmatrix}
        0 & 0 & \frac{k_{\mathrm{off}}^{R}}{u_{1} + s} \\
        \frac{k_{\mathrm{J}}}{u_{2} + s} & 0 & 0
    \end{pmatrix},
    & \quad &\boldsymbol{d}^{\ast}(s) =&
    \begin{pmatrix}
        0 & \frac{k_{\mathrm{off}}^{R}}{u_{1} + s} \\
        \frac{k_{\mathrm{J}}}{u_{2} + s} & 0
    \end{pmatrix}.
\end{IEEEeqnarray*}
We evaluate \eqref{eq:Anderson-Laplace} by noting that $\sum_{k=0}^{\infty} (\boldsymbol{d}^{\ast}(s))^k = (\one + \boldsymbol{d}^{\ast}(s))\sum_{k=0}^{\infty} (\boldsymbol{d}^{\ast}(s))^{2k} = (\one + \boldsymbol{d}^{\ast}(s)) \frac{(u_{1}+s)(u_{2}+s)}{(u_{1}+s) (u_{2}+s) - k_1 k_2} = (\one + \boldsymbol{d}^{\ast}(s)) \frac{(u_{1}+s)(u_{2}+s)}{(w_{1}+s) (w_{2}+s)}$, where $w_{1}, w_{2}$ are the roots of $s \mapsto w(s) = (u_{1}-s) (u_{2}-s) - k_1 k_2$. Then
\begin{equation}
    \subsmkd^{\ast}(s) = \frac{1}{w(s)} \begin{pmatrix}
        k_{1} k_{\mathrm{J}} & 0 & \frac{k_{\mathrm{off}}^{R} (k_{1} k_{\mathrm{J}}+w(s))}{u_{1} + s} \\
        0 & 0 & \frac{k_{\mathrm{off}}^{R} (k_{1} k_{\mathrm{J}}+w(s))}{u_{1} + s} \\
        0 & \frac{k_{\mathrm{on}} w(s)}{k_{\mathrm{on}} + s} & 0
    \end{pmatrix}.
\end{equation}
The inverse Laplace transform for rational functions is found, e.g., in \cite{polyanin2008handbook} (Sec. 7.2.2.). As the coarse graining $\coarse \colon \substsp \to \{\mathrm{J}, \mathrm{ON}, \mathrm{OFF} \}$ (defined as the projection to the first coordinate in $\substsp$ and renaming) is injective, the recurrent hazard function of state $\mathrm{J}$ is given directly by \eqref{eq:MrP-hazard-function}. Thus we obtain an explicit expression for $\lambda^{\IN{} \OUT{}}$, with initial condition that $\OUT{}$ had an arrival at $t=0$. In a similar fashion it has been shown that the holding time $\tau$ of the marginal renewal process $\OUT{}$ has Laplace transformed probability density\cite{hakkinen2016characterizing}
\begin{equation}
   f_{\tau}^{\ast}(s) =  \frac{k_{1} k_{\mathrm{J}} (k_{\mathrm{on}}+s)}{(r_{1} + s)(r_{2} + s)(r_{3} + s)},
\end{equation}
where $r_{1}< r_{2}<r_{3}$ are the roots of $s \mapsto (s-k_{\mathrm{on}}) k_{1} k_{2} + ((k_{\mathrm{on}}-s)(u_{1}-s) k_{\mathrm{on}} k_{\mathrm{off}}^{R})(u_{2}-s)$. Again the Laplace transform is a rational function, such that by \cite{polyanin2008handbook} (Sec. 7.2.2.)
\begin{equation}
    f_{\tau}(t) = k_{1} k_{\mathrm{J}} \sum_{i=1}^{3} \frac{k_{\mathrm{on}} - r_{i}}{\prod_{j \neq i} (r_{j} - r_{i})} e^{- r_{i} t}.
\end{equation}
With the quantities derived so far, we can express the MI $\I(\IN{[0,T]}; \OUT{[0,T]} \mid \dd \OUT{0}=1)$ via Proposition \ref{prop:phi-renewal-density-evolution} and the MIR $\IR(\IN{}; \OUT{})$ via the expressions in Appendix \ref{subsec:MIR-3-state}. We now turn to the system $(\INcat,\OUT{})$.

According to  Sec. \ref{sec:EvalMI} the filtered augmented sMk of the marginal system representing $\OUT{}(\INcat)$ can be written as a block-diagonal matrix $\SMK(t) = \mathrm{diag}(\elSMK{\incat}{\incat}(t) \colon \incat \in \mathcal{\INcat} )$, where
\begin{equation}
    \elsmkd{\incat}{\incat}(t) = f_{\tau}(t) |_{k_{\mathrm{off}}^{R}= k_{\mathrm{off}} (\incat [R]_{0} +(1 - \incat)[R]_{1})}.
\end{equation}
Since $\OUT{}= N(\mathrm{J}_0) + N(\mathrm{J}_1)$ our coarse graining is $\coarse(\mathrm{J}_{\incat})=\mathrm{J}$ for all $\incat \in \mathcal{\INcat}$, such that $\OUT{}= N(\mathrm{J})$. By \eqref{eq:SuffStat0-Cconditional} we have $(\SuffStat{0}^{\INcat, \OUT{}})_{\incat^{\prime}} = \Prob(\subembMC{0} = \mathrm{J}_{\incat^{\prime}} \mid \INcat=\incat) = \delta_{\incat^{\prime} \incat}$, where $\incat^{\prime} \in \mathcal{\INcat}$, and by \eqref{eq:SuffStat0-C} we obtain $(\SuffStat{0}^{\OUT{}})_{\incat^{\prime}} = \Prob(\subembMC{0} = \mathrm{J}_{\incat^{\prime}} ) = \Prob(\INcat = \incat^{\prime})$.
\begin{figure}[t]
    \centering
    \includegraphics[width=\columnwidth]{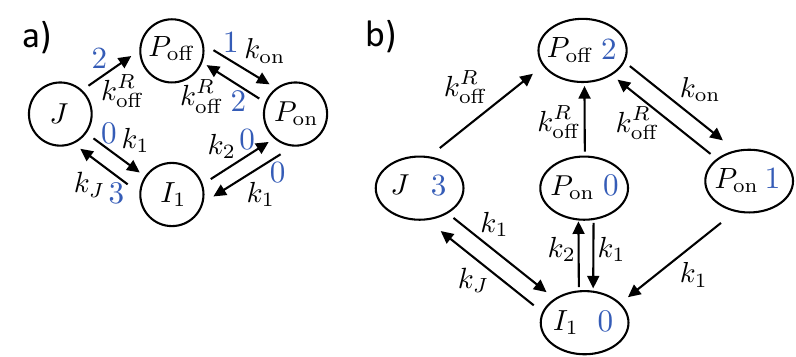}
    \caption{Visualization of the state space augmentation for the case study; a) shows the transition graph of the model of interest with transition class indices (blue) assigned to the transitions; b) depicts the corresponding augmented state space.}
    \label{fig:Filter-Case-Study}
\end{figure}
}{}
\end{document}